\documentclass[reprint,aps,twocolumn,pre,nofootinbib]{revtex4-1}
\usepackage{verbatim,amsmath,amssymb,graphicx,placeins}
\usepackage[margin=0.5 in]{geometry}
\usepackage{xcolor}
\usepackage{slashed}

\newcommand{\ceq}{c^{\text{eq}}}
\newcommand{\ueq}{C^{\text{eq}}}
\newcommand{\Lad}{L_{\text{ad}}}
\newcommand{\Ld}{L_\text{ev}}
\newcommand{\Lp}{L_{\text p}}
\newcommand{\mx}{{\boldsymbol x}}
\newcommand{\my}{{\boldsymbol y}}
\newcommand{\mz}{{\boldsymbol z}}
\newcommand{\nv}{{\boldsymbol\nu}}

\newcommand{\mts}{\widehat{\mathcal S}}
\newcommand{\mtd}{\widehat{\mathcal D}}
\newcommand{\mtk}{\widehat{\mathcal K}}
\newcommand{\mti}{\widehat{\mathcal I}}

\newcommand{\mA}{{\boldsymbol {\mathcal A}}}
\newcommand{\mB}{{\boldsymbol {\mathcal B}}}
\newcommand{\mD}{{\boldsymbol {\mathcal D}}}

\newcommand{\msd}{\widehat{\slashed{\mathcal D}}}

\newcommand{\bwt}{{\bar w}}

\newcommand{\Lef}{L_{\text{eff}}}

\begin{document}

\title{Emergence of local geometric laws of step flow in homoepitaxial growth}
\author{Ian Johnson}
\email{ijohnso6@umd.edu}
	\affiliation{Department of Mathematics, University of Maryland, College Park, Maryland 20742, USA}
	\author{Dionisios Margetis}
	\email{diom@umd.edu}
	\affiliation{Institute for Physical Science and Technology, and Department of Mathematics, and Center for Scientific Computation and Mathematical Modeling, University of Maryland, College Park, Maryland 20742, USA}
	\date{\today}

\begin{abstract}
Below the roughening transition, crystal surfaces exhibit nanoscale line defects, steps, that move by exchanging atoms with their environment. In homoepitaxy, we analytically show how the motion of a step train \color{black} in vacuum under \emph{strong desorption} can be approximately described by nonlinear laws that depend on local geometric features such as the curvature of each step, as well as suitably defined effective terrace widths. \color{black} We assume that each step edge, a free boundary, can be represented by a smooth curve in a fixed reference plane for sufficiently long times. \color{black} Besides surface diffusion and evaporation, the processes under consideration include kinetic step-step interactions in slowly varying geometries, \color{black} material deposition on the surface from above, attachment and detachment of atoms at steps, step edge diffusion, and step permeability. Our methodology relies on \emph{boundary integral equations} for the adatom fluxes responsible for step flow. By applying asymptotics, which effectively treat the diffusive term of the free boundary problem as a singular perturbation, we describe an intimate connection of universal character between step kinetics and local geometry. 
\end{abstract}
\maketitle

\section{Introduction}
\label{sec:Intro}

Epitaxial growth comprises a multitude of kinetic processes and thermodynamic effects such as atom hopping and elastic effects on crystal surfaces~\cite{PimpinelliVillain1999,MichelyKrug2004,Misbahetal2010}. In homoepitaxial growth, in particular, the deposited material is the same as the one of the substrate, in contrast to heteroepitaxy. At temperatures below the roughening transition, the crystal surface morphological evolution at the nanoscale is driven by the motion of line defects which resemble steps and have atomic height~\cite{Misbahetal2010,JeongWilliams1999}. This step flow regime is evident in numerous experimental observations of crystal growth in vacuum or in solution (see, e.g.,~\cite{Wangetal2016,DeYoreo2009,Sazakietal2012,Asakawa2016,Shtukenbergetal2013,Shtukenbergetal2015,Chernov2005,Higgins2000,Tengetal1998, Paloczietal1998}). The reliable description of step dynamics is essential in the predictive modeling of nanostructure evolution, with applications that span microelectronics, energy storage, catalysis and drug design. 

A widely known theory of step motion is the Burton-Cabrera-Frank (BCF) model~\cite{BCF}; see also the earlier works by Kossel and Stranski~\cite{Kossel1927,Stranski1928} and an important extension by Chernov~\cite{Chernov1961}. The BCF model has been successfully applied to many epitaxial phenomena~\cite{JeongWilliams1999,QiuOrme2008}, including relaxation and coarsening~\cite{Laietal2019,Tanakaetal1997,IsraeliKandel1999,Margetis2007,Johnson2019}, bunching instabilities~\cite{Misbahetal2010,Krug_in_Voigt,Stoyanov1991,Foketal2007}, stochastic nanoscale fluctuations~\cite{Misbahetal2010,Ghez1993,OPLMisbah1998a,OPLMisbah1998b,LiuMetiu1994,Patrone2010,Patrone2011} and evolution of crystal facets~\cite{IsraeliKandel1999,Margetisetal2006,Foketal2008,MargetisNakamura2011,Liuetal2019}. By the BCF theory, each step moves by exchanging atoms with its environment as adsorbed atoms (adatoms) diffuse on the adjacent terraces. The projection of the step edge onto a fixed crystal plane of reference is viewed as a \emph{free boundary}. In surface relaxation, the energy of the whole step configuration decreases with time~\cite{PimpinelliVillain1999,MichelyKrug2004}. This picture has been enriched with step free-energy anisotropy, material deposition from above, evaporation, step edge diffusion and step permeability; for reviews, see~\cite{PimpinelliVillain1999,MichelyKrug2004, Misbahetal2010,JeongWilliams1999}. Notably, the normal step velocity, $v_\perp$, at every point of the step free boundary is dictated by mass conservation; $v_\perp$ is proportional to the total mass flux into the step. Because of adatom diffusion on terraces, the step velocity at each point thus depends on the entire step configuration. In this paper, we show how in homoepitaxy $v_\perp$ can be  expressed in terms of \emph{local geometric features} of the step curve under certain physically motivated assumptions.

Geometric models for the motion of \color{black} free boundaries are not uncommon, and are physically transparent and computationally appealing~\cite{CabreraVermilyea1958,Rashkovich-book,Chernov1987,GiraoKohn1994,SchulzeKohn1999,LeeThorp,LeeThorp20}. Such equations are usually speculated via thermodynamics and mass conservation. Regarding the step flow regime, the connection of geometric motion laws to the BCF model~\cite{BCF} is largely unexplored. If the density as well as the mobility of kinks along the step edge are high enough then the normal step velocity $v_\perp$ is allowed to be pointwise regulated by a geometric Gibbs-Thomson-type relation; see, e.g.,~\cite{Tengetal1998,LeeThorp}. The plausible emergence of such a view in the two-dimensional (2D) setting from the BCF theory, by which the diffusion of adatoms on terraces couples $v_\perp$ to the global geometry, is the subject of our study here. 

In this paper, we analytically derive simplified, geometric-type laws for the motion of a step train \color{black} in vacuum by use of a BCF-type model in 2D. Our analysis indicates how the competition of adatom diffusion and evaporation can dramatically affect the form of the step velocity law. We obtain effective parameters that enter this law in the limit of strong desorption, in the presence of several other kinetic processes. The emergence of such parameters as an asymptotic limit of the BCF theory has apparently not been described before. 

We assume that the step curves are smooth and the step geometry is slowly varying for long enough times. \color{black} We also posit that the desorption rate, $\tau^{-1}$, is sufficiently large so that the associated diffusion length, $\Ld=\sqrt{D_s\tau}$, is small compared to the linear size and radius of curvature of the step and the widths of the neighboring terraces, where $D_s$ is the terrace diffusivity.  We employ an asymptotic method that is justified by the length scale separation of this system. 

In particular, we show how a geometric-type step velocity law can emerge as an asymptotic limit from  terrace diffusion, desorption, and atom attachment and detachment at the step edge, in the step configuration. Our result illustrates how this velocity is coupled to effective widths of adjacent terraces. \color{black} In the special case of a single step, our finding reduces to a version of motion by curvature. \color{black} We enrich this asymptotic result with kinetic effects such as the Ehrlich-Schwoebel barrier~\cite{Ehrlich,Schwoebel}, step permeability~\cite{Zangwill1992}, and step edge diffusion~\cite{OPL2001,Krug_in_Voigt}. 

Our approach relies on the conversion of the BCF-type motion laws to a system of \emph{boundary integral equations} for the adatom mass fluxes perpendicular to the two sides of each step edge. An ingredient of this formalism is Green's function for terrace diffusion with desorption, in the quasi-steady approximation. Because of desorption, Green's function decays with a length scale equal to $\Ld$. Integral formalisms for epitaxy can also be found in~\cite{OPLMisbah1998a,Huangetal06}, but their underlying settings involve the full diffusion equation (in spacetime) with objectives different from ours. In these works, as well as in this paper, the spatial nonlocality due to surface diffusion is captured by the boundary integral terms. In~\cite{Huangetal06}, however, desorption is not considered. On the other hand, in~\cite{OPLMisbah1998a} desorption is taken into account through the suitable time scale of the associated propagator, while the strong-desorption limit (which is of interest here) is not studied. 

The role of evaporation in the kinetics of stepped surfaces has been pointed out by BCF~\cite{BCF}. By explicitly solving a version of their model, these authors demonstrated that the lateral speed of a single circular step becomes linear with curvature for sufficiently strong desorption~\cite{BCF}. Since then, step motion laws of similar character in more complicated geometries are often speculated physically, yet without direct recourse to desorption; see, e.g.,~\cite{Rashkovich-book}. 

\begin{figure}[h]
\includegraphics[scale=0.275,trim=0.1in 0.5in 0in 0.5in]{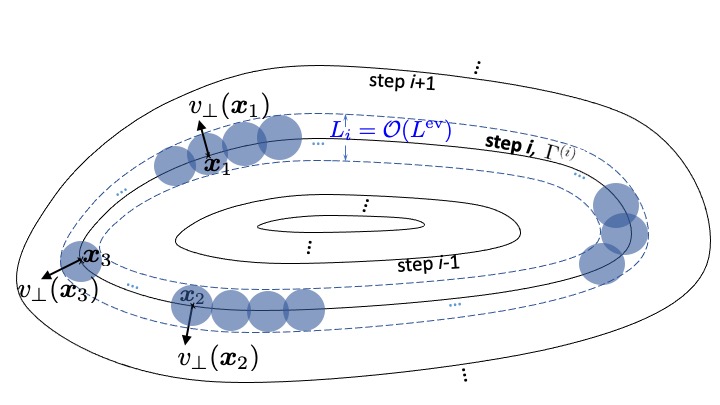}
\centering{}\caption{Schematic of geometry and the role of strong desorption. Solid curves: Step edges, $\Gamma^{(i)}$ (top view); $i=1,\,2,\,\ldots$. Interior of dashed curves: Boundary layer of width $L_i$ for adatom diffusion along entire step $i$; $L_i$ is of the order of $\Ld=\sqrt{D_s\tau}$. This layer is formed by much smaller regions, shown as circular disks of centers $\mx$ on step $i$, e.g., $\mx= \mx_1,\,\mx_2,\,\mx_3$. The normal step velocity $v_\perp(\mx)=v_{i,\perp}(\mx)$ of step $i$ is affected by parameters of kinetic boundary conditions at this step in each respective disk.}
\label{fig:Geometry}
\end{figure}

For strong desorption, the diffusion length $\Ld$ roughly expresses the width of a curved strip in the terraces adjacent to the step as a \emph{boundary layer} of adatom diffusion; see Fig~\ref{fig:Geometry}. The adatom density varies appreciably in the direction normal to the step inside this layer, according to the boundary condition of atom attachment and detachment at the step edge. Away from this layer, the adatom density approaches some constant value fixed by the material deposition from above, except for points close to another step. In this vein, for each point of the step edge, $\Ld$ defines the linear size of a ``domain of influence'' (circular disk in Fig.~\ref{fig:Geometry}) for the local step velocity. The smallness of this domain in comparison to the linear size and radius of curvature of the step shape and the widths of the neighboring terraces, enables the reduction of the BCF-type equations to a geometric motion law. The step velocity determined in this way is only affected by parameters of the boundary condition on the step curve inside this domain, in the vicinity of the respective step edge point.

More precisely, the full BCF-type step velocity equals~\cite{JeongWilliams1999}
\begin{equation*}
v_\perp(\mx)=M \left[\bar c(\mx)-\left(e^{\mu_s(\mx)/T}-1\right)\right]
\end{equation*}
at every point $\mx$ of the step curve, in the absence of step edge diffusion.
In this relation, $\bar c(\mx)$ is the adatom supersaturation at the step edge, defined as a suitable kinetic average of the local adatom density relative to an equilibrium concentration of a straight step; \color{black}  $M$ is the step mobility; $\mu_s(\mx)$ is the step chemical potential which comes from the variation of the total step free energy; and $T$ is the absolute temperature (in units with  $k_B=1$). The supersaturation $\bar c(\mx)$ is determined from solving the adatom diffusion equation, which includes desorption, on terraces; thus, $\bar c(\mx)$ should depend on the whole geometry. Our analysis shows how, for strong enough desorption and sufficiently wide terraces, $v_\perp(\mx)$ approximately reduces to a form that only depends on $\mu_s(\mx)$ and combinations of kinetic lengths and effective terrace widths. 

Hence, a  highlight of our approach is the replacement  of the adatom supersaturation at the step by a simple expression involving the step chemical potential via asymptotics. A geometric law for the step velocity emerges if the step chemical potential is dominated by step stiffness. The parameters of this effective description are obtained explicitly, and can be useful in the modeling of step flow in various settings. \color{black}

Our treatment points to open problems. The presence or formation of corners in the step curve cannot be treated by our asymptotics. Another interesting issue is the effect of stochastic step fluctuations. \color{black} In the case of heteroepitaxy, not addressed here, one might expect that even in the strong desorption limit of that setting the step velocity law would retain a nonlocal term due to long-range elasticity  (see, e.g.,~\cite{Dondl2016}). Since we invoke elements of the BCF theory, we do not directly address the connection of geometric motion laws for steps to the atomistic dynamics on the lattice. Crystal growth in aqueous solutions lies beyond our scope. 

The remainder of this paper is organized as follows. In Sec.~\ref{sec:BCF} we review the BCF theory, particularly the joint effect of adatom diffusion and desorption. In Sec.~\ref{sec:radial}, we focus on the case with strong desorption for concentric circular steps, as an extension of the BCF study~\cite{BCF}. Section~\ref{sec:main} introduces the boundary integral formalism of step flow, and the derivation of a basic asymptotic formula for the step velocity in 2D. This formula accounts for kinetic step-step interactions. In Sec.~\ref{sec:extensions}, we provide extensions of these results to include step edge diffusion and step permeability. Section~\ref{sec:numerics} presents numerical simulations for validation of our method, and discussion of predictions and limitations. In Sec.~\ref{sec:conclusion}, we conclude the paper with a summary of results. 
 
\section{Review of BCF model}
\label{sec:BCF}

In this section, we review elements of the BCF model by including desorption and step permeability. We assume that the terraces between steps are much wider than the diffusion length $\Ld$. This setting favors the localization of the terrace adatom density and flux near each step edge. Aspects of this localization are outlined via boundary layer theory in the end of this section; see also Sec.~\ref{sec:radial}.

We note in passing that, from an atomistic view, the BCF theory relies on the diluteness of the adatom system on the crystal lattice. Hence, a necessary condition for using the BCF model in our treatment is that the P\'eclet number $Pe=FA^2/D_s$ is small, where $A$ is the atomic area~\cite{Schneideretal2018}.  

\subsection{Adatom diffusion and step energy}
\label{subsec:diffusion}

We consider a monotone step train in 2D. For a top view of the configuration, see Fig.~\ref{fig:Geometry}. The projections of the  terraces on a fixed crystallographic plane, say, the $xy$ plane, are the regions 
$\Omega^{(i)}$ where $i=0,\,1,\,\ldots N$; $\Omega^{(N)}$ is unbounded and corresponds to the material substrate. The terrace region $\Omega^{(i)}$ is bounded by the smooth step curves $\Gamma^{(i)}$ and $\Gamma^{(i+1)}$; $\Gamma^{(0)}$ reduces to the origin and $\Gamma^{(N+1)}$ denotes a curve approaching infinity. For definiteness, suppose that the steps are descending with increasing $i$. The total number of the actual steps is not necessarily conserved. We assume that this number is an arbitrary constant $N$ in the time interval of interest.\color{black}

Let $c_i(\mx,t)$ denote the density of adatoms on the $i$-th terrace, $\Omega^{(i)}$, at time $t$. 
This $c_i(\mx,t)$ satisfies 
\begin{equation}
\frac{\partial c_i}{\partial t} = F + D_s \left(\Delta - \Ld^{-2} \right)c_i~, \quad \mx\ \mbox{in}\ \Omega^{(i)}~. \label{eq:bcf_diff}
\end{equation}
In the above, $D_s$ is the terrace diffusion constant, $F$ is the deposition flux (number of atoms per unit area per time), and the term proportional to $\Ld^{-2}$ amounts to evaporation or desorption with constant rate $\tau^{-1}$; $\Ld=\sqrt{D_s\tau}$ and $\Delta$ denotes the 2D Laplacian. We assume that the adatom diffusion is isotropic, and atoms are deposited on the surface from above at a constant rate.

In the spirit of BCF~\cite{BCF}, we employ the quasi-steady approximation for the concentration field $c_i(\mx,t)$. Accordingly, we set $\partial c_i/\partial t\simeq 0$ in Eq.~\eqref{eq:bcf_diff}, and determine the step velocity through the adatom fluxes at the step. In other words, we assume that terrace diffusion is fast compared to other kinetic processes. In this vein, the solution of the diffusion equation on each terrace is replaced by a steady state. At each point $\mx$, the density $c_i(\mx,t)$ evolves with time only through the location of steps at time $t$. Hence, the velocity of each step is determined by the instantaneous geometry of all steps (and not its history). For ease of notation, we will suppress the time ($t$-) dependence of $c_i$ and related variables.

Equation~\eqref{eq:bcf_diff} is supplemented with suitable boundary conditions, which account for kinetic processes at the steps. BCF require that the adatom density have a local equilibrium value at the step edge (Dirichlet condition)~\cite{BCF}. Typical extensions of this condition dictate that the adatom flux normal to each side of the step curve be linear with the respective limiting values of the adatom concentration. Hence, at the steps (labeled by index $j=i,\,i+1$) bounding the $i$th terrace we impose the Robin-type  conditions~\cite{Chernov1961,Ehrlich,Schwoebel,Zangwill1992,PimpinelliVillain1999}
\begin{subequations}\label{eqs:BCF-conds}
\begin{align}
\pm \nv^{(j)}(\mx)\cdot \nabla c_i({\mx}) \cdot  & = \frac{1}{\Lad^\pm} \left [c_i(\mx) - \ceq_j(\mx) \right]\nonumber\\ 
&\pm \frac{1}{\Lp} \left[c_j(\mx) - c_{j-1}(\mx) \right]~, 
\label{eq:bcf_bc_general}
\end{align}
where $\mx$ lies in curve $\Gamma^{(j)}$; $j=i$ ($+$ sign) or $j=i+1$ ($-$ sign),  and $\nv^{(j)}$ is the unit normal vector on $\Gamma^{(j)}$ that points toward lower terraces, outward from the whole structure. \color{black} The left-hand side of Eq.~\eqref{eq:bcf_bc_general} displays a quantity equal to $1/D_s$ times the adatom flux normal to $\Gamma^{(j)}$ outward from terrace $\Omega^{(i)}$. On the right-hand side of Eq.~\eqref{eq:bcf_bc_general}, the first term expresses the deviation of $c_i(\mx)$ from the local equilibrium adatom density, $\ceq_j(\mx)$; while the second term accounts for step permeability~\cite{Zangwill1992}. The quantity $\Lad^\pm$ denotes the (kinetic) attachment-detachment length  $\Lad^\pm=D_s/k^\pm$ where $k^\pm$ is the kinetic parameter with units of velocity for atom exchange between a step edge and the lower ($+$) or upper ($-$) terrace. The asymmetry of this exchange expresses the Ehrlich-Schwoebel barrier~\cite{Ehrlich,Schwoebel}. For a positive Ehrlich-Schwoebel barrier, we have
$k^+>k^-$ and thus $\Lad^+<\Lad^-$. We include step permeability via the length $\Lp=D_s/k_{\text p}$, where $k_{\text p}$ is a kinetic parameter  for the direct hopping of atoms from the vicinity of a step to the adjacent terrace. 

Because the outermost terrace, $\Omega^{(N)}$, is an unbounded region we need to include a boundary condition for the adatom density as $|\mx|\to \infty$. This condition accounts for the balance, or equilibration, between deposition and desorption, viz.,
\begin{equation}
\lim_{|\mx| \to \infty} c_N(\mx)=F\Ld^2/D_s=F\tau\quad (\mx\ \mbox{in}\ \Omega^{(N)})~. \label{eq:bcf_bc_infty}
\end{equation}
\end{subequations}
More generally, the adatom density $c_i(\mx)$ with $i<N$ should approach this limit away from steps, if the width of the terrace $\Omega^{(i)}$ is much larger than $\Ld$; see Sec.~\ref{subsec:bl}.

Next, we describe the velocity law of the free boundary. By mass conservation, the (normal) $i$-th step velocity $v_{i,\perp}(\mx)$ in the direction of $\nv^{(i)}(\mx)$ on curve $\Gamma^{(i)}$ is driven by the total flux of adatoms from the neighboring terraces to the step. In the absence of step edge diffusion, $v_{i,\perp}(\mx)$ is given by~\cite{BCF}
\begin{equation*}
v_{i,\perp}(\mx) = D_sA \left \{\nabla c_i(\mx) - \nabla c_{i-1}(\mx) \right\} \cdot \nv^{(i)}(\mx) 
\end{equation*}
where $\mx$ lies in $\Gamma^{(i)}$. Recall that $A$ denotes the atomic area.

At this stage, we need to specify the local equilibrium adatom concentration $\ceq_i$ which enters Eq.~\eqref{eq:bcf_bc_general}. This quantity expresses  thermodynamic effects, which may include the step stiffness as well as elastic-dipole and entropic repulsive interactions between steps~\cite{PimpinelliVillain1999,JeongWilliams1999}. 
By invoking the Gibbs-Thomson relation~\cite{JeongWilliams1999}, we write (in units with $k_B=1$)
\begin{subequations}
\begin{equation}\label{eq:ceq-BCF}
\ceq_i(\mx)=c_s \exp \Biggl( \frac{\mu_i(\mx)}{T} \Biggr)~,\quad \mx\ \mbox{in}\ \Gamma^{(i)}~.
\end{equation}
Here, $\mu_i(\mx)$ is the chemical potential of the $i$-th step and $c_s$ is the (fixed) equilibrium adatom density of an isolated straight step. The step chemical potential, $\mu_i(\mx)$, a thermodynamic force, is given by the variation with respect to the step shape of the total step free energy, $E_{\mathrm{st}}$, which  depends on the overall geometry of the system (at any given time $t$). For example, suppose that the step curve $\Gamma^{(i)}$ can be described by $r=r_i(\theta)$ in polar coordinates $(r,\theta)$ with $-\pi<\theta\le \pi$. Accordingly, $\mu=\mu_i$ equals~\cite{Krug_in_Voigt}
\begin{equation}\label{eq:mu-var-BCF}
\mu=A\frac{\delta E_{\mathrm{st}}}{\delta r}~,
\end{equation}
\end{subequations}
where all step curves other than $r=r_i$ are frozen.
This formula expresses the variational derivative of $E_{\mathrm{st}}$ with respect to the polar-distance function $r=r_i(\theta)$. Therefore, in principle $\mu=\mu_i$ is a function of the polar angle, $\theta$. \color{black} If the contribution of the step stiffness, $\tilde\gamma$, which comes from the line tension of step $i$, dominates in $\mu_i$ then $\mu_i\simeq \tilde\gamma \kappa_i$ where $\kappa_i$ is the (local) step curvature. Our analysis in this paper does not rely on the precise dependence of $\ceq_i$ on $\mx$. In  Sec.~\ref{subsec:validity}, however, we use a particular choice of $\tilde\gamma$ in order to carry out numerical simulations and validate our approach.  

In addition, we shift the adatom concentration field, $c_i(\mx)$, by a constant in order to transform the BCF-type equations into a form independent of the deposition flux, $F$. Recall that $Pe=FA^2/D_s\ll 1$. We define the variables
\begin{align*}
C_i(\mx) =c_i(\mx)-F\tau~,\quad \ueq_i(\mx) = \ceq_i(\mx)-F\tau~,\quad \mbox{all}\ i~,
\end{align*}
which leaves invariant the adatom flux, $\boldsymbol J_i=-D_s \nabla c_i$. The governing equations for the shifted concentration $C_i(\mx)$ read
\begin{subequations} \label{eqs:bcf_adj_system}
\begin{align}
\Delta C_i(\mx) &= \Ld^{-2} C_i~,\quad \mx\ \mbox{in}\ \Omega^{(i)}~, \label{eq:bcf_adj_pde} \\
\pm \nv^{(j)}(\mx)\cdot &\nabla C_i(\mx)  = \frac{1}{\Lad^\pm} \left[C_i(\mx) - \ueq_j(\mx)\right]  \nonumber \\
& \pm \frac{1}{\Lp} \left[C_j(\mx) - C_{j-1}(\mx)\right]~,\ \mx\ \mbox{in}\ \Gamma^{(j)}~, \label{eq:bcf_adj_bc} \\
&\lim_{|\mx| \to \infty} C_N(\mx) = 0~. \label{eq:CN-lim}
\end{align}
In Eq.~\eqref{eq:bcf_adj_bc}, we set $j=i$ ($+$ sign) or $j=i+1$ ($-$ sign), in correspondence to Eq.~\eqref{eq:bcf_bc_general}. Of course, $C_i(\mx)$ must be bounded. We recognize Eq.~\eqref{eq:bcf_adj_pde} as the modified Helmholtz equation. The form of the step velocity law in the transformed variables remains intact, viz.,
\begin{equation}
v_{i,\perp}(\mx) = D_sA \left \{\nabla C_i(\mx) - \nabla C_{i-1}(\mx) \right\} \cdot \nv^{(i)}(\mx)~. \label{eq:bcf_gen_velocity} 
\end{equation}
\end{subequations}

We now comment on the validity of the quasi-steady approximation from the perspective of continuous adatom diffusion. This simplification is expected to hold for sufficiently long times, $t$. In particular, we are interested in the regime where $t\gg \tau$ and the length $\Ld=\sqrt{D_s \tau}$ is small compared to the linear size of the step curve. Regarding the external deposition flux $F$, the P\'eclet number $Pe=F A^2/D_s$ must be small (as mentioned above)~\cite{Schneideretal2018}. We should also add the assumption that $e^{-t/\tau}F\tau$, which signifies the effect of deposition in the time domain, is small compared to typical adatom concentration values on either side of the step edge. 

\subsection{Strong desorption and scale separation}
\label{subsec:bl}

Next, we delineate the role of strong desorption by direct recourse to the system of Eqs.~\eqref{eq:bcf_adj_pde}--\eqref{eq:bcf_gen_velocity}. Let us neglect step permeability for simplicity, taking $\Lp=\infty$.

For strong desorption, it is tempting to directly take the limit $\Ld\to 0$ ($\tau\to 0$) in Eq.~\eqref{eq:bcf_adj_pde}. The naive approach of eliminating the diffusive term ($\Delta C_i$) everywhere in the $i$th terrace would not allow $C_i$ to satisfy the boundary conditions at the bounding steps, labeled by  $i$ and $i+1$; cf. Eq.~\eqref{eq:bcf_adj_bc}. This observation calls for treating the diffusive term as a \emph{singular perturbation} of the (free) boundary value problem~\cite{Hinch-book}.

Therefore, the enforcement of the boundary conditions for atom attachment and detachment upon the shifted adatom density $C_i(\mx)$ motivates the use of relatively thin boundary layers around the step edges. Each layer lies in the vicinity of the whole step curve, and has a width of the order of $\Ld$ (see Fig.~\ref{fig:Geometry}). Let us briefly consider local curvilinear coordinates in the directions perpendicular and tangential to a step edge. Adopting the language of boundary layer theory~\cite{Hinch-book}, we can assert that $C_i(\mx)$ changes rapidly, at the scale of $\Ld$, in the perpendicular direction but varies slowly in the tangential direction inside the \emph{inner region}. This density decays to zero in the outer region.

This view suggests that the normal step velocity $v_{i,\perp}(\mx)$ of the $i$-th step at point $\mx$ is only affected by parameters such as the shifted equilibrium adatom concentration $\ueq_i(\my)$ of the boundary condition for atom attachment and detachment at points $\my$ of the step in the vicinity of $\mx$. Consequently, a geometric motion law for the step can emerge if the step chemical potential $\mu_i(\mx)$ has a dominant contribution from the step stiffness; see also Sec.~\ref{sec:radial}.

We will describe this reduction via asymptotics on integral equations for the adatom flux normal to steps.  The alternate approach of applying boundary layer theory, or separation of the spatial variables into fast and slow~\cite{DMKohn06}, to the free boundary problem for the diffusion equation on terraces is feasible but lies beyond the scope of this paper (see Sec.~\ref{sssec:electrom}).

\section{The paradigm of radial geometry} 
\label{sec:radial}

In this section, we study the geometry with concentric circular steps as an example of how step motion can be approximately reduced to local geometric laws. The radial setting is prototypical since it allows us to explicitly solve the multi-step boundary value problem of adatom diffusion formulated in Sec.~\ref{subsec:diffusion}, which forms an extension of the one-step case studied in~\cite{BCF}. We formally apply asymptotics to the explicit solution by assuming that the diffusion length $\Ld$ is small enough. Our leading-order asymptotic result has a generic form, which offers insight into the more general 2D setting. We also discuss the idea of the shifted adatom density localization near step edges as a way of motivating the boundary integral formalism of Sec.~\ref{sec:main}. For simplicity, in this section we neglect step permeability, taking $\Lp= \infty$ in Eq.~\eqref{eq:bcf_adj_bc}. Step permeability is studied in some detail in Sec.~\ref{subsec:transp}.

\subsection{Explicit solution and asymptotics}
\label{subsec:explicit-radial}
Consider the setting in which the step curves $\Gamma^{(i)}$ are concentric circles with center at the origin and radii $r_i(t)$ ($i=1,\,2,\,\ldots,\,N$); cf. Fig.~\ref{fig:Geometry}. The non-extremal $i$-th terrace $\Omega^{(i)}$ is the annulus bounded by circles $\Gamma^{(i)}$ and $\Gamma^{(i+1)}$ for $i=1,\,2,\,\ldots,\,N-1$.  The extremal terrace $\Omega^{(0)}$ is the circular disk of radius $r_1(t)$ while the terrace $\Omega^{(N)}$ is unbounded, containing the points at distance $r>r_N(t)$ from the origin. Similar formulations without desorption can be found in~\cite{IsraeliKandel1999,UwahaWatanabe2000,MAS05}. Notably, equations of motion for circular steps under desorption are described in~\cite{DM-Nakamura2013}, with emphasis on connections between simplified versions of these equations and their full continuum limits.\color{black}

First, we explicitly solve the boundary value problem implied by Eqs.~\eqref{eq:bcf_adj_pde}--\eqref{eq:CN-lim}, in regard to the modified Helmholtz equation for the shifted concentration $C_i(\mx)$. We seek a rotationally symmetric solution for $C_i(\mx)$, which we denote $C_i(r)$ by abusing notation. This density satisfies the problem  
\begin{align*}
\frac{1}{r} \partial_r(r \partial_r C_i) & = \Ld^{-2} C_i~,\quad r_i(t)<r<r_{i+1}(t)~,\\
\pm \partial_r C_i(r)\big|_{r=r_j} & = \frac{1}{\Lad^\pm} \big[C_i(r_j) - \ueq_j\big] 
\end{align*}
for $j=i$ ($+$ sign) and $j=i+1$ ($-$ sign) where $i=0,\,1,\,\ldots,\, N$. Here, we take $r_0=0$ and $r_{N+1}=\infty$;  and use the symbol $\partial_r=\partial/\partial r$.  Regarding the terraces $\Omega^{(0)}$ and $\Omega^{(N)}$, the same boundary condition is applied at $r=r_1$ (for $\Omega^{(0)}$, with the $-$ sign), and at $r=r_N$ (for $\Omega^{(N)}$, with the $+$ sign). In addition, the density $C_0(r)$ must be bounded at the origin while $C_N(r)\to 0$ as $r\to \infty$.

By solving the equation for $C_i(r)$,  we find ($i=0,\,\ldots,\,N$)
\begin{equation*}
C_i(r)=a_i\, I_0(r/\Ld) + b_i\, K_0(r/\Ld)~,\quad r_i(t)<r<r_{i+1}(t)~,
\end{equation*}
where $I_n$ and $K_n$ are the $n$-th order modified Bessel functions of the first and second kind, respectively~\cite{Bateman-II}. We need to take $b_0=0$ and $a_N=0$, since $K_0(R)\to \infty$ (logarithmically) as $R\to 0$ and $I_0(R)\to \infty$ (exponentially) as $R\to\infty$.

The coefficients $a_i$ and $b_i$ can be determined explicitly by use of the atom attachment and detachment (Robin-type) condition at the steps $\Gamma^{(i)}$ and $\Gamma^{(i+1)}$ which bound the terrace $\Omega^{(i)}$. After some algebra, we obtain ($i=1,\,2,\,\ldots,\,N-1$)
\begin{subequations}
\begin{align}
a_i = & \frac{1}{\Lambda_i} \left[\left ( \frac{K_1(R_i)}{\Ld} + \frac{K_0(R_i)}{\Lad^+} \right ) \frac{\ueq_{i+1}}{\Lad^-} \right. \nonumber \\
& \left. + \left ( \frac{K_1(R_{i+1})}{\Ld} - \frac{K_0(R_{i+1})}{\Lad^-} \right ) \frac{\ueq_i}{\Lad^+}\right]~, \label{eq:ai_exact} \\
b_i = & \frac{1}{\Lambda_i} \left[\left ( \frac{I_1(R_{i+1})}{\Ld} + \frac{I_0(R_{i+1})}{\Lad^-} \right ) \frac{\ueq_i}{\Lad^+} \right. \nonumber \\
&  \left. +\left ( \frac{I_1(R_i)}{\Ld} - \frac{I_0(R_i)}{\Lad^+} \right ) \frac{\ueq_{i+1}}{\Lad^-}\right]\quad (i\neq 0,\,N)~. \label{eq:bi_exact}
\end{align}
Furthermore, for $i=0$, we have $b_0=0$ and 
\begin{align}
a_0=\left[\frac{I_0(R_1)}{\Lad^-}+\frac{I_1(R_1)}{\Ld}\right]^{-1}\frac{\ueq_1}{\Lad^-}~; \label{eq:a0_exact}
\end{align}
while, for $i=N$, we have $a_N=0$ and
\begin{align}
	b_N=\left[\frac{K_0(R_N)}{\Lad^+}+\frac{K_1(R_N)}{\Ld}\right]^{-1}\frac{\ueq_N}{\Lad^+}~. \label{eq:bN_exact}
\end{align}
\end{subequations}
In the above, we introduce the nondimensional step radii
$R_i=r_i/\Ld$ ($i=1,\,2,\,\ldots,\,N$), and also define the quantities
\begin{align*}
\Lambda_i = & \left ( \frac{K_1(R_i)}{\Ld} + \frac{K_0(R_i)}{\Lad^+} \right ) \left ( \frac{I_1(R_{i+1})}{\Ld} + \frac{I_0(R_{i+1})}{\Lad^-} \right ) \nonumber \\
& - \left ( \frac{I_1(R_i)}{\Ld} - \frac{I_0(R_i)}{\Lad^+} \right ) \left ( \frac{K_1(R_{i+1})}{\Ld} - \frac{K_0(R_{i+1})}{\Lad^-} \right )~,
\end{align*}
for $i=1,\,\ldots,\, N-1$.

To determine the step velocities in terms of the step radii, we should compute the total (radial) adatom flux into each curve $\Gamma^{(i)}$. This flux is defined by $J_{i}^{\rm tot}=-D_s [\partial_r C_{i-1}(r)-\partial_r C_{i}(r)]$ at $r=r_i$, and is given by the formula ($i=1,\,\ldots,\,N$)
\begin{equation}\label{eq:Ji-tot}
J_{i}^{\rm tot}=\frac{D_s}{\Ld} \left[(a_i-a_{i-1}) I_1(R_i) + (b_{i-1}-b_i) K_1(R_i) \right]~. 
\end{equation}
The $i$-th step velocity in the radial direction is $v_{i,\perp}=A J_{i}^{\rm tot}$.

By Eq.~\eqref{eq:Ji-tot}, $J_i^{\rm tot}$ is a sum of contributions each of which is proportional to $\ueq_j$ for $j=i,\,i\pm 1$. Hence, we can write
\begin{equation*}
J_i^{\rm tot}=\frac{D_s}{\Ld}\left(A_i \ueq_i +B_{i+} \ueq_{i+1}+B_{i-} \ueq_{i-1}  \right);\ i=2,\,\ldots,\, N-1.	
\end{equation*}
The coefficients $A_i$ and $B_{i\pm}$ can be explicitly expressed in terms of step radii; see Appendix~\ref{app:radial} for a matrix formalism. \color{black} We omit the respective exact formulas for $A_i$ and $B_{i\pm}$ here. 

Next, we focus on step configurations in which  all terrace widths are large compared to $\Ld$, viz., $r_i - r_{i-1} \gg \Ld$ for $i=1,\,2,\,\ldots,\,N$ (where $r_0=0$). We seek the leading-order asymptotic formula for each step velocity $v_{i,\perp}$ via the flux $J_i^{\rm tot}$. This task calls for the asymptotic evaluation of the coefficients $A_i$ and $B_{i\pm}$ for $R_{i}-R_{i-1}\gg 1$. 

By invoking the large-argument approximations for $I_n(R)$ and $K_n(R)$, i.e., $I_n(R)\simeq e^R/\sqrt{2\pi R}$ and $K_n(R)\simeq e^{-R}\sqrt{\pi/(2R)}$ as $R\to \infty$~\cite{Bateman-II}, we compute
\begin{align*}
\Lambda_i & = \frac{e^{R_{i+1}-R_i}}{2\sqrt{R_i R_{i+1}}} \left ( \frac{1}{\Ld} + \frac{1}{\Lad^+} \right ) \left ( \frac{1}{\Ld} + \frac{1}{\Lad^-} \right ) \left\{1+o(1)\right\}
\end{align*}
if $R_{i+1}-R_i \gg 1$ for all $i$. Here, the the symbol $o(1)$ accounts for neglected terms which involve negative powers of $R_{i+1}$ and $R_i$; $i=1,\,\ldots,\,N-1$. Notably, the approximate formula for $\Lambda_i$ is invariant under the interchange of $\Lad^+$ and $\Lad^-$. Similarly, we obtain approximate formulas for $A_i$ and $B_{i\pm}$.

The substitution of these asymptotic formulas into the expression for the total adatom flux $J_{i}^{\rm tot}$ in turn furnishes 
\begin{align}
v_{i,\perp}\simeq &  \frac{2AD_s \Ld}{(\Ld+\Lad^+)(\Ld+\Lad^-)} \left\{-\left(1+\frac{\Lad^+ +\Lad^-}{2\Ld} \right)\, \ueq_i \right. \nonumber \\
& +e^{-(R_{i+1}-R_i)} \sqrt{\frac{R_{i+1}}{R_i}} \, \ueq_{i+1} \nonumber \\
& \left. + e^{-(R_i-R_{i-1})} \sqrt{\frac{R_{i-1}}{R_i}}  \,\ueq_{i-1}\right\}~,\label{eq:radial_flux_asym}
\end{align}
if $i=2,\,\ldots,\,N-1$.  Note that the contributions of $\ueq_{i\pm 1}$ are exponentially small in this formula, since $|R_{i\pm 1}-R_i|\gg 1$. Equation~\eqref{eq:radial_flux_asym} can be extended to the remaining steps $\Gamma^{(i)}$, namely, the curve $\Gamma^{(1)}$ (for $i=1$) and curve $\Gamma^{(N)}$ ($i=N$). We omit the resulting expressions here.

A special case is the geometry with a single, isolated circular step. The   (radial) velocity of this step becomes
\begin{equation*}
	v_\perp  \simeq  -AD_s \left ( \frac{1}{\Ld+\Lad^-} + \frac{1}{\Ld+\Lad^+}  \right )\, \ueq~,
\end{equation*}
where $\ueq=c_s \exp(\mu/T)-F\tau$ and the step chemical potential, $\mu$, comes from the variation of the isotropic step free energy. The linearization of this exponential for $|\mu|\ll T$ with $\Lad^+=\Lad^-\ll \Ld$ yields a formula consistent with the result by BCF~\cite{BCF}. In particular, suppose that $\mu$ is dominated by the step stiffness, $\tilde\gamma$, and $F\tau> c_s$. The step velocity is thus reduced to the form \color{black} 
\begin{equation*}
	v_\perp  \simeq v_{\infty} \left(1-\frac{r_{\text c}}{r}\right)~,
\end{equation*}
where $v_\infty$ denotes the velocity of an isolated straight step and $r_{\text c}$ is the radius of a ``critical nucleus''~\cite{BCF}. In our setting, we must define
\begin{equation*}
	v_\infty=-AD_s\left(\frac{1}{\Lad^+ +\Ld}+\frac{1}{\Lad^-+\Ld} \right)(c_s-F\tau)
\end{equation*}
and 
\begin{equation*}
	r_{\text c}=\frac{c_{s}}{F\tau -c_s} \frac{\tilde\gamma}{T}~.
\end{equation*}
Hence, in this limit $v_\perp$ is linear with the step curvature. 

A few further remarks on Eq.~\eqref{eq:radial_flux_asym} are in order. First, for each $i$ the terms proportional to $\ueq_j$ with $j=i\pm 1$ describe kinetic step-step interactions. Second, in  many situations of interest the condition $|R_{i\pm 1}-R_i|\ll R_i$ holds for some $i$; thus, the corresponding factors $\sqrt{R_{i\pm 1}/R_i}$ can be replaced by unity. Third, the coefficients of $\ueq_j$ for $j=i$ and $j=i\pm 1$ were computed to the leading order in the scaled terrace widths $|R_{i\pm 1}-R_i|$. In fact, for the coefficient of $\ueq_i$  we neglected terms that involve negative powers of $R_i$ and $R_{i\pm 1}$ which, although small compared to the leading-order term of this coefficient, can be much larger than the displayed coefficients for $\ueq_{i\pm 1}$. Thus, our asymptotic formula is viewed as a formal description of distinct physical contributions of equilibrium densities to the step velocity.

A generic feature of Eq.~\eqref{eq:radial_flux_asym} unfolds. Specifically, the contributions of $\ueq_{i \pm 1}$ to step velocity $v_{i,\perp}$ decay exponentially with $|r_{i \pm 1}-r_i|/\Ld$. This behavior is indicative of the effect of the diffusion boundary layer in the vicinity of each step (Sec.~\ref{subsec:bl}); the steps $i\pm 1$ lie in the outer region of step $i$. Suppose that elastic-dipole and other step-step interactions are neglected in the step chemical potential $\mu_i$, which controls $\ueq_i$. Consequently, we explicitly verify that steps separated by terraces (circular annuli) that are large compared to $\Ld$ are decoupled in their motion because of the effect of strong desorption. The emerging step velocity is linear in the curvature (inverse radius) of the step edge.



\subsection{Localization via desorption in non-radial geometry}
\label{subsec:localization}

At the risk of redundancy we now repeat the idea about the role of strong desorption in the general 2D setting. Our discussion  motivates the boundary integral formalism (Sec.~\ref{sec:main}).

In the non-radial case, the diffusion equation of the BCF model cannot be solved exactly even in the quasi-steady limit. However, key physical aspects of desorption underlying the analysis of the radial geometry persist in the more general 2D setting. Specifically, the diffusive flux connecting any two points separated by a distance much larger than $\Ld$ on each terrace is negligible.  Hence, the dominant contribution to the step velocity at any given point $\mx$ of the step curve comes from a neighborhood of $\mx$ that has linear size comparable to the diffusion length $\Ld$ (cf. Fig.~\ref{fig:Geometry}). 

This localization of the adatom concentration, and normal adatom flux,  at the step is intimately related to the presence of a boundary layer in the sense of Sec.~\ref{subsec:bl}.  Our goal is to derive a step velocity law analogous to Eq.~\eqref{eq:radial_flux_asym} in the non-radial setting by exploiting this property.  

To this end, we will employ a method that directly extracts information \emph{only} about the fluxes normal to the step edges. This method is described in Sec.~\ref{sec:main}. The starting point is the exact conversion of Eqs.~\eqref{eq:bcf_adj_pde}--\eqref{eq:CN-lim} into a system of boundary integral equations for the shifted adatom density on each side of every step edge. 

The joint effect of terrace diffusion and desorption is expressed through the kernel in the boundary integral equations. It is worthwhile to describe this kernel, denoted by $G(\mx,\my)$. This $G$ comes from the fundamental solution, or Green's function, of Eq.~\eqref{eq:bcf_adj_pde} in the plane (infinite terrace); and is rotationally symmetric in $\mx-\my$. If we define $G$ via the equation $\{\Delta_{\mx}-\Ld^{-2}\}G(\mx,\my)=\delta(\mx-\my)$, where $\Delta_{\mx}$ is the Laplacian in $\mx$, and resort to Bessel functions, we find \color{black} 
\begin{equation}\label{eq:Green-func}
G(\mx,\my)=-\frac{1}{2\pi} K_0 \left ( \frac{|\mx-\my|}{\Ld} \right )
\end{equation}
for all points $\mx$, $\my$ with $\mx\neq \my$. Recall that $K_0(z)$ is the zeroth-order modified Bessel function of the second kind~\cite{Bateman-II}. 
Evidently, $\sqrt{|\mx -\my|}G(\mx,\my)$ decays exponentially with the scaled distance $|\mx-\my|/\Ld$ for $|\mx-\my|\gg \Ld$~\cite{Bateman-II}. 

\section{Boundary integral formalism and asymptotics} 
\label{sec:main}

In this section, our task is twofold. First, we exactly convert the BCF-type free boundary problem of Sec.~\ref{sec:BCF} for the shifted adatom concentration field on the terraces into a system of boundary integral equations for the adatom flux normal to steps. An advantage of this formalism is that it circumvents the need to compute the adatom concentration and flux in the terraces. Second, we apply asymptotic methods to the ensuing boundary integral equations when the diffusion length $\Ld$ is sufficiently small (thus, desorption is strong). We are able to obtain analytical expressions for the adatom fluxes normal to step edges. This approach allows us to derive a local geometric law for the step velocity. The interested reader may find our main result for the step velocity in Sec.~\ref{subsec:emergence}, skipping the related derivations.

We posit that the step curves are smooth for long enough times, in the time interval of interest~\cite{Note-facets}. In our formalism, we allow for step energy anisotropy; hence, the free energy of each step may depend on the step orientation in the fixed crystallographic plane of reference. This anisotropy is assumed to be compatible with our step curve smoothness hypothesis. In addition, we set $\Lp=\infty$ in Eq.~\eqref{eq:bcf_adj_bc}, neglecting permeability. We will include this effect in Sec.~\ref{subsec:transp}.

\subsection{General integral formalism} 
\label{subsec:formalism}
In this subsection, we derive boundary integral equations along the step curves by using the BCF model~\cite{BCF} and elements of potential theory~\cite{Courant-book,Kellogg-book}. Consider the $i$-th terrace $\Omega^{(i)}$ which is bounded by curves $\Gamma^{(i)}$ and $\Gamma^{(i+1)}$ ($i=1,\,\ldots,\,N$). To simplify notation, let $C(\mx)=C_i(\mx)$ denote the shifted adatom density in the fixed terrace $\Omega^{(i)}$, suppressing the terrace index for the density.

\begin{figure}[h]
\includegraphics[scale=0.25,trim=0.1in 0.5in 0in 0.5in]{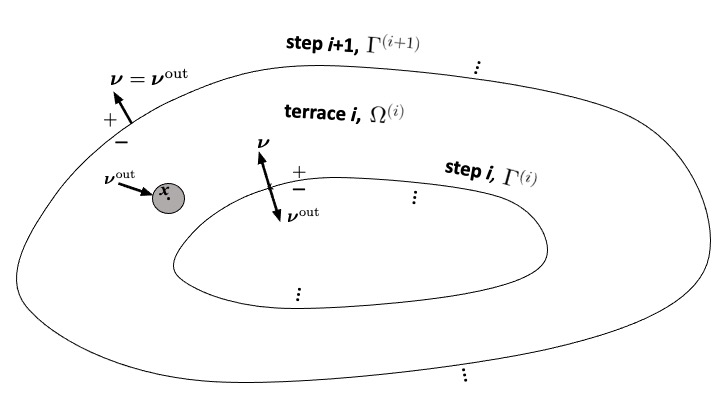}
\centering{}\caption{Geometry for derivation of boundary integral equations. For the $i$-th terrace ($\Omega^{(i)}$), the fixed observation point $\mx$ is excluded from integration by removal of a small circular disk (grey shaded region) centered at $\mx$. The unit normal $\nv$ along the terrace boundary, consisting of curves $\Gamma^{(i)}$ and $\Gamma^{(i+1)}$, points outward from the whole structure. The unit normal $\nv^{\text{out}}$ points outward from the punctured $i$-th terrace. The $\pm$ signs on each side of a step indicate the convention, relative to $\nv$, adopted in Eqs.~\eqref{eqs:ies_for_omegai_no_bc} and~\eqref{eqs:ies_for_omegai_bc}.}
\label{fig:Green-geom}
\end{figure}

First, we derive an integral representation for the shifted density $C$ in the $i$-th terrace $\Omega^{(i)}$ in terms of the values of $C$ and its normal derivative at the bounding steps via Green's function $G$, Eq.~\eqref{eq:Green-func}~\cite{Kellogg-book,Courant-book,Evans-book}. Following the standard approach, let us  
fix a point $\mx$ in $\Omega^{(i)}$, multiply both sides of Eq.~\eqref{eq:bcf_adj_pde}
by $G(\mx,\my)$, 
and suitably integrate over a ``punctured region'' $\tilde{\Omega}^{(i)}$ which comes from the terrace $\Omega^{(i)}$ by removal of a small circular disk 
centered at $\mx$; see Fig.~\ref{fig:Green-geom}. Integration by parts in the resulting equation yields the line integral
\begin{equation*} 
\int_{\partial \tilde{\Omega}^{(i)}} \left\{G(\mx,\my) \frac{\partial C(\my)}{\partial \nu^{\rm out}}
-  \frac{\partial G(\mx,\my)}{\partial \nu^{\rm out}} C(\my)\right\} \, ds_\my =0~,
\end{equation*}
which is defined along the boundary ($\partial \tilde{\Omega}^{(i)}$) of  $\tilde{\Omega}^{(i)}$. Hence, the line integral is carried out along the smooth curves $\Gamma^{(i)},\Gamma^{(i+1)}$ and 
the small circle centered at $\mx$. Note that  
$\partial Q/\partial \nu^{\rm out}$ ($Q=C,\,G$) denotes the derivative of function $Q$ along the boundary of $\tilde{\Omega}^{(i)}$ in the direction of the unit normal vector $\nv^{\text{out}}$ that points outward from $\tilde{\Omega}^{(i)}$.  In the line integral, the functions $C$ and $\partial C/\partial \nu^{\rm out}$, evaluated at the point $\my$ of the boundary, are the limits of $C(\mz)$ and $\nv^{\text{out}}(\my)\cdot \nabla C(\mz)$, respectively, as $\mz$ approaches $\my$ from the $i$-th terrace. 
By contracting the small circle to the point $\mx$, we obtain~\cite{Kellogg-book,Evans-book}
\begin{equation*}
\int_{\partial \Omega^{(i)}} \left\{G(\mx,\my) \frac{\partial C(\my)}{\partial \nu^{\rm out}} 
-  \frac{\partial G(\mx,\my)}{\partial \nu^{\rm out}} C(\my)\right\} \, ds_\my = -C(\mx)~. \label{eq:green_rep_ibp_mu}
\end{equation*}
The integration path is the boundary of terrace $\Omega^{(i)}$, which consists solely of curves $\Gamma^{(i)}$ and $\Gamma^{(i+1)}$. 

We now introduce operator notation for later algebraic convenience. To this end, for each step curve we use the unit normal vector $\nv$ pointing outward from the whole structure; see Fig.~\ref{fig:Green-geom}. Hence, we write $\partial Q/\partial\nu^{\text{out}}= \partial Q/\partial\nu$ on $\Gamma^{(i+1)}$ and $\partial Q/\partial\nu^{\text{out}}= -\partial Q/\partial\nu$ on $\Gamma^{(i)}$ (for $Q=C,\,G$). Accordingly, the equation for $C(\mx)$ is recast to the form 
\begin{align}\label{eq:Green-form}
&\mts_{i+1}\biggl[\frac{\partial C}{\partial\nu}\biggr](\mx) 
- \mts_i\biggl[\frac{\partial C}{\partial \nu}\biggr](\mx) \nonumber\\
& -\mtd_{i+1}[C](\mx) + \mtd_i[C](\mx) = -C(\mx) 
\end{align}
where $\mx$ lies in $\Omega^{(i)}$.  By adopting the formalism of potential theory~\cite{Kellogg-book}, we recognize  $\mts_j$ and $\mtd_j$ ($j=i,\,i+1$) as \emph{single- and double-layer potential} operators, respectively, along step curve $\Gamma^{(j)}$. For a physically admissible density or normal-flux  function $f$ on step curve $\Gamma^{(j)}$, these operators are defined via
\begin{subequations} \label{eqs:layer_potential_defns}
\begin{align}
\mts_j[f](\mx) & = \int_{\Gamma^{(j)}} G(\mx,\my) f(\my)\, ds_\my~, \label{eq:defn_S} \\
\mtd_j[g](\mx) & = \int_{\Gamma^{(j)}} \frac{\partial G(\mx,\my)}{\partial \nu(\my)} g(\my)\, ds_\my
\label{eq:defn_D}
\end{align}
where $\mx$ may lie anywhere, with the \emph{exception} of curve $\Gamma^{(j)}$ in Eq.~\eqref{eq:defn_D}. In Eq.~\eqref{eq:Green-form}, the operators $\mts_j$ and $\mtd_j$ act on functions $f$ and $g$ identified with the boundary values of $\nv\cdot \nabla C(\mz)$ and $C(\mz)$, respectively, as $\mz$ approaches the curve $\Gamma^{(j)}$ from \emph{inside} $\Omega^{(i)}$. These one-sided limits are implied by the notation for the shifted density, $C=C_i$. Evidently, the single- and double-layer potentials in Eq.~\eqref{eqs:layer_potential_defns} with $f=\nu\cdot \nabla C$ and $g=C$ have the dimension of the adatom concentration. \color{black} 

We should add a few comments. First, recall that terraces $\Omega^{(i)}$ are labeled by $i=0,\,1,\,\ldots,\,N$. Equation~\eqref{eq:Green-form} is derived for points $\mx$ in non-extremal terraces, if $1\le i\le N-1$. This description can be extended to extremal terraces $\Omega^{(i)}$ (for $i=0,\,N$) via the convention that $\mts_j$ and $\mtd_j$ are zero for $j=0,\,N$. Second, the adatom density and its normal derivative can be discontinuous across step edges. The jump in the adatom flux normal to a step is needed for a nonzero step velocity. The one-sided shifted adatom density is related to the respective normal derivative via Robin-type condition~\eqref{eq:bcf_adj_bc}. Third, the integrand of the double-layer potential $\mtd_j[g](\mx)$, seen in Eq.~\eqref{eq:defn_D}, has a singularity in $\my$ as $\mx$ approaches $\Gamma^{(j)}$.

To obtain the desired boundary integral equations, we need to separate the contribution of the aforementioned singularity.
Hence, in Eq.~\eqref{eq:Green-form} we let $\mx$ approach any point $\tilde \mx$ of step curve $\Gamma^{(j)}$ ($j=i,\,i+1$) from terrace $\Omega^{(i)}$. We invoke the limit~\cite{Kellogg-book}
\begin{equation*} 
\lim_{\mx \to \tilde \mx} \mtd_j[g](\mx) = \mp \frac{1}{2} g(\tilde \mx) + \msd_j[g](\tilde \mx)~,\quad \tilde \mx\ \mbox{in}\ \Gamma^{(j)}~,
\end{equation*}
for $j=i$ ($-$ sign) or $j=i+1$ ($+$ sign)~\cite{Comment-sign}. The first term of the limit represents the contribution of the singularity. In the second term, we introduce the operator $\msd_j$ defined by
\begin{equation}
\msd_j[g](\tilde \mx)= \int_{\Gamma^{(j)}} \frac{\partial G(\tilde\mx,\my)}{\partial \nu(\my)} g(\my)\,
ds_\my~, \quad \tilde\mx\ \mbox{in}\ \Gamma^{(j)}~. \label{eq:defn_K}
\end{equation}
\end{subequations}
This integral is well defined along a smooth curve $\Gamma^{(j)}$.  

Thus, Eq.~\eqref{eq:Green-form} yields two distinct relations, depending on the step curve in which $\tilde\mx$ lies. Dropping the tilde, we obtain
\begin{subequations} \label{eqs:ies_for_omegai_no_bc}
\begin{align}
& \mts_{i+1}\biggl[\biggl(\frac{\partial C}{\partial \nu}\biggr)^- \biggr](\mx) - \mts_i\biggl[\biggl(\frac{\partial C}{\partial \nu}\biggr)^+\biggr](\mx)\nonumber \\
& -  \mtd_{i+1}[C^-](\mx)+ \msd_i[C^+](\mx) = -\frac{1}{2} C^+(\mx)
\end{align}
for $\mx$ lying in curve $\Gamma^{(i)}$; and
\begin{align}
& \mts_{i+1}\biggl[\biggl(\frac{\partial C}{\partial \nu}\biggr)^- \biggr](\mx) - \mts_i\biggl[\biggl(\frac{\partial C}{\partial \nu}\biggr)^+\biggr](\mx)\nonumber \\
& -  \msd_{i+1}[C^-](\mx)+ \mtd_i[C^+](\mx) = -\frac{1}{2} C^-(\mx) 
\end{align}
\end{subequations}
for $\mx$ in curve $\Gamma^{(i+1)}$. The symbol $Q^\pm$ ($Q=C,\,\partial C/\partial\nu$) for the shifted density and its normal derivative along a step edge denotes the boundary value of $Q$ on each side of the step edge   
\emph{relative to} the unit normal $\nv$ (see Fig.~\ref{fig:Green-geom}). Recall that the vector $\nv$ points outward from the whole structure.
Although we currently focus on the steps bounding terrace $\Omega^{(i)}$, our choice of notation will be 
useful later, when we compute the step velocity by an asymptotic method. 

Equation~\eqref{eqs:ies_for_omegai_no_bc} is not in the desired form as yet, since both the shifted adatom density and normal flux are used. However, the step velocity is driven by the total flux into the step edge. Hence, it is advantageous to eliminate $C^\pm$ by use of Robin-type boundary conditions~\eqref{eq:bcf_adj_bc}. After some algebra, we recast Eq.~\eqref{eqs:ies_for_omegai_no_bc} to the following relations:
\begin{subequations} \label{eqs:ies_for_omegai_bc}
	\begin{align} \label{eq:ies_for_omegai_bc-1}
		& \left(\mts_{i+1} + \Lad^- \mtd_{i+1}\right)\biggl[\biggl(\frac{\partial C}{\partial \nu}\biggr)^-\biggr](\mx) \nonumber \\
		& +  \left(-\mts_i + \textstyle{\frac{1}{2}} \Lad^+ \mti + \Lad^+ \msd_i\right)\biggl[\biggl(\frac{\partial C}{\partial \nu}\biggr)^+\biggr](\mx) \nonumber \\
		& = \left(-\textstyle{\frac{1}{2}}\mti -\msd_i \right )[\ueq_i](\mx)  +  \mtd_{i+1}[\ueq_{i+1}](\mx)  
	\end{align}
for points $\mx$ in $\Gamma^{(i)}$; and
\begin{align} \label{eq:ies_for_omegai_bc-2}
	& \left ( \mts_{i+1} - \textstyle{\frac{1}{2}} \Lad^- \mti+ \Lad^- \msd_{i+1}\right )\biggl[\biggr(\frac{\partial C}{\partial \nu}\biggr)^-\biggr](\breve\mx)  \nonumber \\
	& + \left ( -\mts_i + \Lad^+ \mtd_i \right )\biggl[\biggr(\frac{\partial C}{\partial \nu}\biggr)^+\biggr](\breve\mx)  \nonumber \\
	& = \left ( -\textstyle{\frac{1}{2}} \mti + \msd_{i+1} \right )[\ueq_{i+1}](\breve\mx) -\mtd_i[\ueq_i](\breve\mx)  
\end{align}
\end{subequations}
for $\breve\mx$ in $\Gamma^{(i+1)}$.  In the above, $\mti$ is the identity operator. In these equations, the unknown functions (of $\my$) are the normal derivatives $(\partial C/\partial\nu)^{\pm}$ on the sides of the steps bounding the $i$-th terrace. These functions are integrated along step curves $\Gamma^{(i)}$ and $\Gamma^{(i+1)}$.

By integral equations~\eqref{eq:ies_for_omegai_bc-1} and~\eqref{eq:ies_for_omegai_bc-2}, one can in principle determine the fluxes from terrace $\Omega^{(i)}$ into the bounding step edges in terms of the equilibrium concentrations $\ueq_i$ and $\ueq_{i+1}$. This formalism is an exact consequence of the BCF-type free boundary problem of Sec.~\ref{sec:BCF} in the quasi-steady approach. Hence, by complementing these  
integral relations with the respective equations for terrace $\Omega^{(i-1)}$, one can obtain the step velocity of the $i$th step edge, $\Gamma^{(i)}$. The local step velocity law will eventually emerge via asymptotics in the limit of sufficiently small $\Ld$.

\subsection{Leading-order asymptotics} 
\label{subsec:asymptotics}

Next, we simplify Eq.~\eqref{eqs:ies_for_omegai_bc} when the diffusion length $\Ld$ is small compared to the linear size and radius of curvature of each step curve. We focus on the leading-order formulas of this limit. The core idea is that for strong  desorption the system is characterized by a length scale separation, namely, the adatom flux along the step  varies slowly in the scale of $\Ld$. The step geometry is also assumed to be slowly varying, which means that step edges bounding a terrace are treated as almost locally parallel to each other. We repeat that the interested reader may directly seek the main result for the step velocity in Sec.~\ref{subsec:emergence}, skipping details of our scheme. \color{black}

\begin{figure}[h]
\includegraphics[scale=0.25,trim=0.8in 0.8in 0in -0.2in]{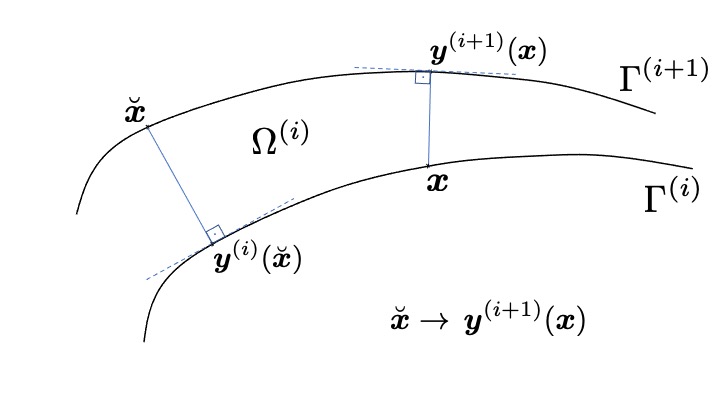}
\centering{}\caption{Schematic on manipulations for Eqs.~\eqref{eqs:ies_first_approx} and~\eqref{eqs:ies_fin_approx}. Points in steps $\Gamma^{(i)}$  and $\Gamma^{(i+1)}$, which bound the $i$th terrace $\Omega^{(i)}$, are mapped to points of minimal distance along $\Gamma^{(i+1)}$ and $\Gamma^{(i)}$, respectively. The point $\mx$ lying in $\Gamma^{(i)}$ is mapped to the point $\my^{(i+1)}(\mx)$ of $\Gamma^{(i+1)}$ that minimizes the distance from $\mx$ to the points of $\Gamma^{(i+1)}$. Similarly, the point $\breve\mx$ of $\Gamma^{(i+1)}$ is mapped to $\my^{(i)}(\breve\mx)$ in $\Gamma^{(i)}$. We let $\breve\mx$ be $\my^{(i+1)}(\mx)$. Each mapping is assumed to be one-to-one.}
\label{fig:Geodesic-geom}
\end{figure}

The ideas of strong desorption and slowly varying step geometry are used for the derivation of asymptotic formulas for the normal derivatives $(\partial C/\partial\nu)^{\pm}$ along a step curve, in the spirit of the radial case (Sec.~\ref{sec:radial}). The step velocity law can be obtained accordingly. The procedure has two main ingredients. First, for fixed yet arbitrary points lying in the step edges bounding the $i$th terrace, in Eq.~\eqref{eqs:ies_for_omegai_bc} we replace the spatially varying functions $(\partial C/\partial\nu)^{\pm}$ and $C_j^{{\rm eq}}$ ($j=i,\,i+1$) by suitably chosen constants in the integrals for the single- and  double-layer potentials. This approximation can be justified, because the integration kernels decay exponentially with the distance $|\mx-\my|$ scaled by the diffusion length $\Ld$. Second, for slowly varying step geometry we develop a scheme  that yields a closed system of equations for $(\partial C/\partial\nu)^{\pm}$. To this end, we invoke the distance of a point from a curve, as shown in Fig.~\ref{fig:Geodesic-geom}, which leads to the notion of the effective terrace width. Our resulting asymptotic formula for the velocity  $v_{i,\perp}$ of the $i$th step forms a nontrivial extension of Eq.~\eqref{eq:radial_flux_asym} of the radial setting.

\subsubsection{Approximation scheme} 
\label{sssec:scheme}
Consider the integrals of Eq.~\eqref{eqs:ies_for_omegai_bc} in regard to the steps bounding the $i$th terrace.  By asymptotics for Laplace-type integrals~\cite{Erdelyi-book}, we need to single out the values $\my_*$ of the integration variable $\my$ (along any step edge) that minimize the distance $|\my-\mx|$ from a given point $\mx$ in the same or a neighboring step. This task calls for distinguishing the following cases. If the observation point $\mx$ and integration variable $\my$ lie in the same step then $\my_*=\mx$. Otherwise, $\my_*$ is the point of the step that minimizes the distance of all points of this step from $\mx$. In this case, $|\my_*-\mx|$ is defined as the distance of $\mx$ from the respective step curve. More generally, for any point $\mx$ of the plane we employ the notation (Fig.~\ref{fig:Geodesic-geom})
\begin{align*}
	\my_*=\my_*(\mx)=\my^{(j)}(\mx)\quad  \mbox{for}\ \my_*\ \mbox{in}\ \Gamma^{(j)}~,
\end{align*}
assuming that $\my_*$ is uniquely defined. In the special case with both $\mx$ and $\my$ lying in $\Gamma^{(j)}$, we must set $\my^{(j)}(\mx)=\mx$. 

Accordingly, in Eq.~\eqref{eqs:ies_for_omegai_bc} we apply the approximation 
\begin{align*}
	\mtk_{j}[f](\mx)&\simeq \mtk_j[1](\mx)\, f(\my_*(\mx))~;\ \mtk=\mts~,\,\mtd~,\,\msd\  (j=i,\,i+1)~.
\end{align*}
Here, $f$ stands for $(\partial C/\partial \nu)^\pm$, $C_i^{{\rm eq}}$ or $C_{i+1}^{{\rm eq}}$. Hence, Eq.~\eqref{eqs:ies_for_omegai_bc} is approximately reduced to the following system:
\begin{subequations}\label{eqs:ies_first_approx}
\begin{align} \label{eq:ies_first_approx_gammai}
& \left(\mts_{i+1}[1](\mx) + \Lad^- \mtd_{i+1}[1](\mx)\right)\, 
\biggl(\frac{\partial C}{\partial \nu}\biggr)^-(\my^{(i+1)}(\mx)) \nonumber \\
		& + \left(-\mts_i[1](\mx) + {\textstyle \frac{1}{2}} \Lad^+ + \Lad^+ \msd_i[1](\mx)\right) \biggl(\frac{\partial C}{\partial \nu}\biggr)^+(\mx) \nonumber \\
		& \simeq  \left(-{\textstyle\frac{1}{2}} - \msd_i[1](\mx) \right ) \ueq_i(\mx) + \{\mtd_{i+1}[1](\mx)\}\, \ueq_{i+1}(\my^{(i+1)}(\mx)) 
	\end{align}
for any point $\mx$ in curve $\Gamma^{(i)}$; and
\begin{align} \label{eq:ies_first_approx_gammaip1}
	& \left(\mts_{i+1}[1](\breve\mx) - {\textstyle \frac{1}{2}} \Lad^- + \Lad^- \msd_{i+1}[1](\breve\mx)\right) \biggl(\frac{\partial C}{\partial \nu}\biggr)^-(\breve\mx) \nonumber \\
	& + \left(-\mts_i[1](\breve\mx) + \Lad^+ \mtd_i[1](\breve\mx)\right) \biggl(\frac{\partial C}{\partial \nu}\biggr)^+(\my^{(i)}(\breve\mx)) \nonumber \\
	& \simeq  \left ( -{\textstyle\frac{1}{2}}  + \msd_{i+1}[1](\breve\mx) \right ) \ueq_{i+1}(\breve\mx) - \{\mtd_i[1](\breve\mx)\}\, \ueq_i(\my^{(i)}(\breve\mx))
\end{align}
\end{subequations}
 for any point $\breve\mx$ in $\Gamma^{(i+1)}$. 
 Equation~\eqref{eqs:ies_first_approx} does not provide a closed system as yet, since the one-sided derivatives $(\partial C/\partial \nu)^\pm$ are evaluated at a total of four points of the two steps. The system of equations appears under-determined. 
  
 To achieve closure, we assume that the step geometry is slowly varying, inspired by the radial case (Sec.~\ref{subsec:explicit-radial}). First, we choose $\breve \mx$ to be  the point $\my_*(\mx)=\my^{(i+1)}(\mx)$, along curve $\Gamma^{(i+1)}$ (Fig.~\ref{fig:Geodesic-geom}). Second, we approximate $\my^{(i)}(\breve x)$ (in $\Gamma^{(i)}$) by
 \begin{equation*}
 	\my^{(i)}(\breve x)=\my^{(i)}(\my^{(i+1)}(\mx))\simeq \mx
 \end{equation*}
in the arguments of $(\partial C/\partial\nu)^+$ and $\ueq_i$ in Eq.~\eqref{eq:ies_first_approx_gammaip1}, since steps bounding a terrace are treated as nearly parallel to each other. This simplifying assumption permeates our analysis.

Accordingly, we obtain a system of linear equations for $(\partial C/\partial \nu)^+$ at point $\mx$ of step curve $\Gamma^{(i)}$,  and $(\partial C/\partial \nu)^-$ at point $\my_*(\mx)=\my^{(i+1)}(\mx)$ of curve $\Gamma^{(i+1)}$. The system reads
\begin{subequations}\label{eqs:ies_fin_approx}
	\begin{align} \label{eq:ies_fin_approx_gammai}
		& \left(\mts_{i+1}[1](\mx) + \Lad^- \mtd_{i+1}[1](\mx)\right)\, \biggl(\frac{\partial C}{\partial \nu}\biggr)^-(\my_*(\mx)) \nonumber \\
		& + \left(-\mts_i[1](\mx) + {\textstyle \frac{1}{2}} \Lad^+ + \Lad^+ \msd_i[1](\mx)\right) \biggl(\frac{\partial C}{\partial \nu}\biggr)^+(\mx) \nonumber \\
		& \simeq \left(-{\textstyle\frac{1}{2}} - \msd_i[1](\mx) \right )\ueq_i(\mx) + \{\mtd_{i+1}[1](\mx)\}\, \ueq_{i+1}(\my_*(\mx))~, 
	\end{align}
\begin{align} \label{eq:ies_fin_approx_gammaip1}
	& \left[\mts_{i+1}[1](\my_*(\mx)) - {\textstyle \frac{1}{2}} \Lad^- + \Lad^- \msd_{i+1}[1](\my_*(\mx))\right]\biggl(\frac{\partial C}{\partial \nu}\biggr)^-(\my_*(\mx))   \nonumber \\
	& + \left[-\mts_i[1](\my_*(\mx)) + \Lad^+ \mtd_i[1](\my_*(\mx))\right] \biggl(\frac{\partial C}{\partial \nu}\biggr)^+(\mx)   \nonumber \\
	& \simeq  \left ( -{\textstyle\frac{1}{2}}  + \msd_{i+1}[1](\my_*) \right ) \ueq_{i+1}(\my_*) - \{\mtd_i[1](\my_*)\}\ueq_i(\mx)~.
\end{align}
\end{subequations}
 This system pertains to terrace $\Omega^{(i)}$. Recall that the velocity of the $i$th step is determined by the total mass flux into the step. Thus, the above equations should be supplemented with their 
 counterparts for $(\partial C/\partial \nu)^+$ at point $\mx$ of step curve $\Gamma^{(i-1)}$, and $(\partial C/\partial \nu)^-$ at point $\my^{(i)}(\mx)$ of curve $\Gamma^{(i)}$.  Notably, in the radial setting, when the step line tension is isotropic and each step curve $\Gamma^{(i)}$ is a circle, Eq.~\eqref{eqs:ies_fin_approx} reduces to the exact result of Sec.~\ref{subsec:explicit-radial}. In this case, the single- and double-layer potential terms can be evaluated by use of modified Bessel functions; see Appendix~\ref{app:radial}. \color{black}

\subsubsection{Simplified formulas for adatom flux} 
\label{sssec:simplified}
Next, we derive the step velocity law through asymptotic formulas for the fluxes normal to steps.
We apply the 
approximations for strong desorption and slowly varying step geometry introduced in Sec.~\ref{sssec:scheme}.

Consider Eq.~\eqref{eqs:ies_fin_approx}, in regard to the steps bounding the $i$th terrace $\Omega^{(i)}$, keeping also in mind its counterpart for terrace $\Omega^{(i-1)}$. 
By analogy with the radial setting (Sec.~\ref{subsec:explicit-radial}), we explicitly solve Eq.~\eqref{eqs:ies_fin_approx} 
by retaining terms that express kinetic interactions between adjacent steps to the leading order for strong desorption. In this sense, we keep terms of the order of $\exp[-w_{j}(\mx)/\Ld]$, where $\mx$ lies in $\Gamma^{(i)}$ and
\begin{equation}\label{eq:eff-terr}
w_{j}(\mx)=|\my^{(j)}(\mx)-\mx|~,\quad j=i\pm 1~.
\end{equation}
The length $w_j(\mx)$ is an effective
terrace width measuring the distance of point $\mx$ on $\Gamma^{(i)}$ from step $j$. Our labeling of effective terrace widths here is algebraically convenient, and differs from the labeling of terraces. We restore the standard labeling in Sec.~\ref{subsec:emergence}.
For fixed step $i$, we neglect terms that scale as $\exp[-l w_{i \pm 1}(\mx)/\Ld]$, $l>1$. The definition of length $w_j(\mx)$ can be extended to any point $\mx$ of the step configuration, where index $j$ refers to a neighboring step.\color{black}

This procedure can be illustrated by the determinant $\mathfrak D$ of the matrix coefficients of system~\eqref{eqs:ies_fin_approx}, viz.,
\begin{align*}
	\mathfrak D=& \{\mts_{i+1}[1](\mx)+\Lad^-\mtd_{i+1}[1](\mx)\}\{-\mts_{i}[1](\my_*)+\Lad^+\mtd_{i}[1](\my_*)\}\notag\\
	&-\left\{\mts_{i+1}[1](\my_*)-{\textstyle\frac{1}{2}}\Lad^-+\Lad^-\msd_{i+1}[1](\my_*)\right\}\notag\\
	&\quad \times\left\{-\mts_{i}[1](\mx)+{\textstyle\frac{1}{2}}\Lad^+ +\Lad^+\msd_{i}[1](\mx)\right\}~.
\end{align*}
Here, $\mx$ lies in step curve $\Gamma^{(i)}$ and $\my_*=\my_*(\mx)=\my^{(i+1)}(\mx)$ is the point of curve $\Gamma^{(i+1)}$ with minimal distance from $\mx$. A key observation is that the term $\mts_j[1](\mz)$ is of the order of $\Ld$ if $\mz$ is in $\Gamma^{(j)}$ but behaves as $\Ld\exp[-w_{j}(\mx)/\Ld]$ otherwise, while $\mtd_j[1](\mz)$ scales as $\exp[-w_{j}(\mx)/\Ld]$; $\mz=\mx,\,\my_*(\mx)$. On the other hand, $\msd_j[1](\mz)$ is of the order of $\Ld$ times the local curvature, as we show below. Accordingly, we approximate 
\begin{align*}
\mathfrak D&\simeq 	-\left\{\mts_{i+1}[1](\my_*)-{\textstyle\frac{1}{2}}\Lad^-+\Lad^-\msd_{i+1}[1](\my_*)\right\}\notag\\
	&\qquad \times\left\{-\mts_{i}[1](\mx)+{\textstyle\frac{1}{2}}\Lad^+ +\Lad^+\msd_{i}[1](\mx)\right\}~,
\end{align*}
neglecting terms of the order of $\exp[-2 w_{i \pm 1}(\mx)/\Ld]$. We solve system~\eqref{eqs:ies_fin_approx} for $(\partial C/\partial\nu)^+$, approximating the remaining determinant in a fashion similar to the calculation for $\mathcal D$. Thus, we find 
\begin{subequations} \label{eqs:fluxes_first_approx}
	\begin{widetext}
	\begin{align}\label{eq:fluxes_first_approx-1}
		\biggl(\frac{\partial C}{\partial \nu}\biggr)^+(\mx) &\simeq  
		\left \{ {\textstyle \frac{1}{2}} \Lad^+ - \mts_i[1](\mx) + \Lad^+ \msd_i[1](\mx) \right \}^{-1} 
		\left\{ -{\textstyle \frac{1}{2}} \Lad^- + \mts_{i+1}[1](\my^{(i+1)}(\mx)) + \Lad^- \msd_{i+1} [1](\my^{(i+1)}(\mx)) \right\}^{-1} \nonumber \\
		& \times \left \{ -\ueq_i(\mx) \left({\textstyle \frac{1}{2}} + \msd_i[1](\mx) \right) 
		\left [ -{\textstyle\frac{1}{2}} \Lad^- + \mts_{i+1}[1](\my^{(i+1)}(\mx)) + \Lad^- \msd_{i+1}[1](\my^{(i+1)}(\mx)) \right] \right. \nonumber \\
		& \left. + \ueq_{i+1}(\my^{(i+1)}(\mx)) \left [ \mts_{i+1}[1](\my^{(i+1)}(\mx))\, \mtd_{i+1}[1](\mx) 
		+ {\textstyle \frac{1}{2}} \mts_{i+1}[1](\mx) - \msd_{i+1}[1](\my^{(i+1)}(\mx))\, \mts_{i+1} [1](\mx) \right] \right\}~, 
	\end{align}
\end{widetext}
where $\mx$ lies in the $i$th step, $\Gamma^{(i)}$; see Fig.~\ref{fig:Green-geom}.

The next task is to find an expression for $(\partial C/\partial \nu)^-$ on the $i$th step. This task can be carried out without much effort by using the formula for $(\partial C/\partial \nu)^-(\my_*)$ from the solution of Eq.~\eqref{eqs:ies_fin_approx} under suitable replacements. To this end, we assume a one-to-one correspondence of every point $\mx$ in a step to the minimal-distance points $\my_*=\my_*(\mx)$ in an adjacent step. Hence, we solve for $(\partial C/\partial \nu)^-$ at point $\my_*$ and then replace $i$ by $i-1$ and $\my_*=\my^{(i)}(\mx)$ by $\mx$; thus, $\my_*$ becomes $\my^{(i-1)}(\mx)$  in our formula eventually.  The result of this manipulation is
\begin{widetext}
\begin{align}\label{eq:fluxes_first_approx-2}
		\biggl(\frac{\partial C}{\partial \nu}\biggr)^-(\mx) &\simeq 
		\left \{ {\textstyle \frac{1}{2}} \Lad^+ - \mts_{i-1}[1](\my^{(i-1)}(\mx)) + \Lad^+ \msd_{i-1} [1](\my^{(i-1)}(\mx)) \right\}^{-1} 
		\left\{ -{\textstyle \frac{1}{2}} \Lad^- + \mts_i[1](\mx) + \Lad^- \msd_i[1](\mx) \right\}^{-1} \nonumber \\
		& \times \left \{ \ueq_i(\mx) \left ( -{\textstyle \frac{1}{2}} + \msd_i[1](\mx) \right ) 
		\left [ {\textstyle \frac{1}{2}} \Lad^+ - \mts_{i-1}[1](\my^{(i-1)}(\mx)) + \Lad^+ \msd_{i-1} [1](\my^{(i-1)}(\mx)) \right ] \right. \nonumber \\
		& \left. -\ueq_{i-1}(\my^{(i-1)}(\mx)) \left [ -\mts_{i-1}[1](\my^{(i-1)}(\mx)) \mtd_{i-1}[1](\mx) + {\textstyle \frac{1}{2}} \mts_{i-1}[1](\mx) + \msd_{i-1}[1](\my^{(i-1)}(\mx)) \mts_{i-1}[1](\mx) \right ]\right\}~,
	\end{align}
\end{widetext}  
\end{subequations}
where $\mx$ lies in $\Gamma^{(i)}$.
By Eqs.~\eqref{eq:fluxes_first_approx-1} and~\eqref{eq:fluxes_first_approx-2}, we can compute the velocity of the $i$th step in terms of single- and double-layer potentials applied to unity; recall  Eq.~\eqref{eq:bcf_gen_velocity}.

\subsubsection{Asymptotics for single- and double-layer potentials} 
\label{sssec:sd-potentials}
Next, we evaluate the single- and double-layer potential contributions in Eq.~\eqref{eqs:fluxes_first_approx}, where the point $\mx$ lies in $\Gamma^{(i)}$; cf. Eq.~\eqref{eqs:layer_potential_defns}. Regarding integration with respect to $\my$ along step edge $\Gamma^{(j)}$ ($j=i,\,i\pm 1$), we employ the signed arclength parametrization of curve $\Gamma^{(j)}$. Thus, we set   
\begin{equation*}
\my=\my_j(\varsigma)~;\quad  \my_j(0)=\left\{\begin{array}{lr} 
\my^{(j)}(\mx)~, & j= i\pm 1 \cr
 \mx~, & j=i 
 \end{array}
 \right.~.
\end{equation*}
In the above, $\varsigma$ is a (dimensionless) signed arclength that is scaled by $\Ld$. This $\varsigma$ ranges from $-L_j/(2\Ld)$ to $L_j/(2\Ld)$ where $L_j$ is the length of curve $\Gamma^{(j)}$. Our goal is to develop asymptotic formulas when $\Ld$ is small compared to the step edge linear size and radius of curvature.

 We start with the potentials in which both the evaluation point $\my_j(0)$ and the integration variable $\my$ lie in the same step edge
$\Gamma^{(j)}$ ($j=i,\,i\pm 1$). For the single-layer potential $\mts_j[1]$, in particular, we need an approximation for the distance
between two points on the step edge as $\varsigma\to 0$. The desired approximation is
$|\my_j(\varsigma)-\my_j(0)| \simeq |\varsigma|\Ld $, by neglect of terms of the order of $|\varsigma|^3$. \color{black} Hence, a change of the  integration variable from $\my$ to $\varsigma(\my)$ yields
\begin{equation*}
\mts_j[1](\my_j(0)) \simeq -\frac{\Ld}{2\pi} \int_{-\frac{L_j}{2\Ld}}^{\frac{L_j}{2\Ld}} K_0 \left ( |\varsigma| \right ) \, d\varsigma~. 
\end{equation*}
We have dropped terms of the order of $\Ld[\Ld \kappa_j(y_j(0)]^2$ where $\kappa_j(\my)$ is the curvature of $\Gamma^{(j)}$ at $\my$. \color{black}
For $L_j/\Ld\gg 1$, we obtain
\begin{subequations}\label{eqs:asympt-potentials-same_step}
\begin{equation} \label{eq:Si_approx}
\mts_j[1](\my_j(0)) \simeq -\frac{\Ld}{2\pi} \int_{-\infty}^\infty K_0(|\varsigma|) \, d\varsigma = -\frac{\Ld}{2}~,
\end{equation}
neglecting terms of the order of $\sqrt{\Ld/L_j} \exp[-L_j/(2\Ld)]$.

In this vein, regarding the double-layer potential, we have
\begin{align*}
\msd_j[1](\my_j(0)) & =\frac{1}{2\pi} \int_{-\frac{L_j}{2\Ld}}^{\frac{L_j}{2\Ld}} d\varsigma\ K_1\biggl( \frac{|\my_j(\varsigma)-\my_j(0)|}{\Ld}\biggr) \nonumber \\
&
\times \frac{(\my_j(\varsigma)-\my_j(0)) \cdot \nu_j(\varsigma)}{|\my_j(\varsigma)-\my_j(0)|}~.
\end{align*}
We approximate $|\my_j(\varsigma)-\my_j(0)|$ in the kernel argument in a way similar to the single-layer potential case. In addition, we need an approximation 
of $(\my_j(\varsigma)-\my_j(0)) \cdot \nu_j(\varsigma)$, 
as $\varsigma$ approaches $0$. The Taylor expansion of this function yields 
\begin{equation*}
	\frac{(\my_j(\varsigma)-\my_j(0)) \cdot \nu_j(\varsigma)}{|\my_j(\varsigma)-\my_j(0)|} \simeq \frac{1}{2} |\varsigma|\,\Ld\kappa_j(\my_j(0)) \qquad \mbox{as}\ \varsigma\to 0~,
\end{equation*}
where the local curvature $\kappa_j$ of step $j$ is evaluated at point $\my_j(0)$.
In the spirit of the derivation of Eq.~\eqref{eq:Si_approx}, we obtain
\begin{align} \label{eq:Ki_approx}
	\msd_j[1](\my_j(0)) & \simeq \frac{\Ld}{2\pi} \int_{-\infty}^\infty |\varsigma| K_1(|\varsigma|) \frac{\kappa_j(y_j(0))}{2}\,d\varsigma \notag\\
	&= \frac{1}{4} \Ld \kappa_j(\my_j(0))~.
\end{align}
\end{subequations}

Next, in regard to Eq.~\eqref{eqs:fluxes_first_approx} let us consider the potentials $\mts_{j\pm 1}[1](\my_j)$ and $\mtd_{j\pm 1}[1](\my_j)$ for which the evaluation point $\my_j=\my_j(0)$ in $\Gamma^{(j)}$ and the respective integration variable $\my$ lie in different steps, separated by a terrace. In Eq.~\eqref{eqs:fluxes_first_approx}, this situation arises for $j=i$. We will address this problem more generally, considering the step $j$ coupled with a neighboring step $k$ ($k=j\pm 1$). Hence, for each $j$ we carry out the integration in $\my=\my_{k}(\varsigma)$ along the curve $\Gamma^{(k)}$ adjacent to $\Gamma^{(j)}$. Recall that $\my_i(0)=\mx$. The effective terrace widths $w_{k}(\my_j(0))$ are involved in this calculation; cf. Eq.~\eqref{eq:eff-terr}. We assume that each $w_{k}$ is larger than $\Ld$. \color{black}


First, we have the approximation
\begin{equation*}
 |\my_k(\varsigma)-\my_j(0)| \simeq \sqrt{w_{k}^2-(-1 \pm w_{k}\, \kappa_{k})\Ld^2\varsigma^2}~,\ k=j\pm 1~,
\end{equation*}
where $w_k=w_k(\my_j(0))$ and $\kappa_k=\kappa_k(\my_j(0))$. 
Note that 
$(1 \mp w_k \kappa_k)>0$, since $|\varsigma|$ is let to become arbitrarily large eventually in our asymptotics. This implies that the terrace width should not exceed the local radius of curvature.
Without further ado, for the single-layer potential we obtain
\begin{subequations}\label{eqs:asympt-potential-diff_step}
\begin{align} \label{eq:Sj_approx}
\mts_k[1](\my_j(0)) &\simeq  -\frac{\Ld}{2\pi} \left ( 1 \mp w_k\kappa_k \right )^{-1/2} \notag\\
&\qquad \times \int_{-\infty}^\infty K_0 \biggl( \sqrt{\frac{w_k^2}{\Ld^2} + \varsigma^2} \biggr)d\varsigma  \nonumber \\
& = -\frac{\Ld}{2} \left ( 1 \mp w_k \kappa_k \right )^{-1/2} e^{-\frac{w_k}{\Ld}}.
\end{align}
The key feature of this formula is the exponential decay with the scaled terrace width, $w_k/\Ld$. This result is consistent with the findings for the radial setting via an exact solution; cf. Eq.~\eqref{eq:radial_flux_asym}. 
The factor $( 1 \mp w_k \kappa_k)^{-1/2}$ here reduces to $\sqrt{R_{i \pm 1}/R_i}$ which appears in the radial case, if $k=i\pm 1$. \color{black} 

Lastly, we consider the terms involving $\mtd_k$, viz.,
\begin{align*}
\mtd_k[1](\my_j(0)) & =\frac{1}{2\pi} \int_{-\frac{L_k}{2\Ld}}^{\frac{L_k}{2\Ld}} d\varsigma\ K_1\biggl( \frac{|\my_k(\varsigma)-\my_j(0)|}{\Ld}\biggr) \nonumber \\
&
\times \frac{(\my_k(\varsigma)-\my_j(0)) \cdot \nu_k(\varsigma)}{|\my_k(\varsigma)-\my_j(0)|}~.
\end{align*}
Thus, we need an approximation for $(\my_k(\varsigma)-\my_j(0)) \cdot \nu_k(\varsigma)$. 
To incorporate the approximations for the $\mts_k$ and $\mtd_k$ terms into the overall result consistently, 
we seek a two-term asymptotic formula for $\mtd_k[1](\my_j(0))$. 
By a Taylor expansion, we find
\begin{align*}
	(\my_k(\varsigma)-\my_j(0)) \cdot \nu_k(\varsigma) 
	\simeq \pm w_k + \frac{1}{2} (1 \mp w_k \kappa_k)\, (\Ld\kappa_k)\Ld \varsigma^2~,
\end{align*}
where the upper (lower) sign corresponds to $k=j+1$ ($j-1$). \color{black}
Consequently, our computation yields 
\begin{align} \label{eq:Dj_approx}
&\mtd_k[1](\my_j(0)) \simeq 
\frac{1}{\pi} \int_{\frac{w_k}{\Ld}}^\infty d\varsigma\, K_1(\varsigma) \left ( \frac{\pm w_k/\Ld}{\sqrt{1 \mp w_k \kappa_k} \sqrt{\varsigma^2-w_k^2/\Ld^2}} \right. \nonumber \\
& \left. \qquad \hphantom{\frac{1}{\pi} \int_{\frac{w_k}{\Ld}}^\infty d\varsigma} + \frac{1}{2}\frac{\Ld\kappa_k \sqrt{\varsigma^2-w_k^2/\Ld^2}}{\sqrt{1 \mp w_k \kappa_k}} \right ) \nonumber \\
& = \frac{1}{2}\left ( \pm 1 + \Ld \kappa_k/2 \right ) (1 \mp w_k \kappa_k)^{-1/2} e^{-\frac{w_k}{\Ld}}~.
\end{align}
\end{subequations}

So far, we replaced the single- and double-layer potentials by simplified formulas. The remaining task is to express the step velocity in terms of the approximate normal fluxes.

\medskip               

\subsection{Emerging step velocity law} 
\label{subsec:emergence}
Next, we combine the ingredients of our approximation in order to express the step velocity in terms of the local curvature. The substitution of the formulas from Eqs.~\eqref{eqs:asympt-potentials-same_step} 
and~\eqref{eqs:asympt-potential-diff_step} into Eq.~\eqref{eqs:fluxes_first_approx}, and the subsequent use of mass conservation statement~\eqref{eq:bcf_gen_velocity}, yield
\begin{subequations}
\begin{widetext}
\begin{align} \label{eq:main_flux_result}
	& v_{i,\perp}\simeq 
	D_s A \left \{ -\ueq_i \left (\frac{1+\frac{1}{2} \Ld \kappa_i}{\Lad^+ + \Ld + \frac{1}{2} \Lad^+ \Ld \kappa_i} 
	+ \frac{1-\frac{1}{2} \Ld \kappa_i}{\Lad^- + \Ld - \frac{1}{2} \Lad^- \Ld \kappa_i} \right ) \right. \nonumber \\
	& + \ueq_{i+1}\,  (1-\bwt_{i} \kappa_{i+1})^{-1/2}  
	\frac{2\Ld}{\left ( \Lad^+ + \Ld + \frac{1}{2} \Lad^+ \Ld \kappa_i \right ) \left (\Lad^- + \Ld + \frac{1}{2} \Lad^- \Ld \kappa_{i+1} \right )}\  e^{-\frac{\bwt_{i}}{\Ld}}\nonumber \\
	& \left.  + \ueq_{i-1}\, (1+\bwt_{i-1} \kappa_{i-1})^{-1/2}   
	\frac{2\Ld}{\left ( \Lad^+ + \Ld + \frac{1}{2} \Lad^+ \Ld \kappa_{i-1}\right ) \left ( \Lad^- + \Ld + \frac{1}{2} \Lad^- \Ld \kappa_i \right )} e^{-\frac{\bwt_{i-1}}{\Ld}}\right \}~.
\end{align}
\end{widetext}
The velocity $v_{i,\perp}$ is evaluated at point $\mx$ of the $i$th step edge. In the above, the labeling of effective terrace widths $\bwt_j$ is the same as the one for terraces $\Omega^{(j)}$. We employ the notation $\bwt_{i}=w_{i+1}(\mx)=|\my^{(i+1)}(\mx)-\mx|$ and  $\bwt_{i-1}=w_{i-1}(\mx)=|\my^{(i-1)}(\mx)-\mx|$ where $\my^{(j)}(\mx)$ is the point on step $j$ with minimal distance from $\mx$ ($j=i\pm 1$). In a similar vein, the local curvature $\kappa_j$ and equilibrium density $\ueq_j$ of the $j$th step are evaluated at the point $y^{(j)}(\mx)$ ($j=i,\,i\pm 1$); recall that $y^{(i)}(\mx)=\mx$. \color{black} 

We consider Eq.~\eqref{eq:main_flux_result} as a highlight of our results. This formula can be readily extended to the extremal steps
$\Gamma^{(1)}$ and $\Gamma^{(N)}$ by removing the $i-1$ and $i+1$ terms, respectively. For an isolated step (if $N=1$), Eq.~\eqref{eq:main_flux_result} readily becomes
\begin{align} \label{eq:main_flux_1step}
	v_{\perp} &\simeq -D_s A \ueq \left \{\frac{1+\frac{1}{2} \Ld \kappa}{\Lad^+ + \Ld + \frac{1}{2} \Lad^+ \Ld \kappa} \right. \notag \\
	& \left. + \frac{1-\frac{1}{2} \Ld \kappa}{\Lad^- + \Ld - \frac{1}{2} \Lad^- \Ld \kappa} \right \}~;
\end{align}
$\kappa=\kappa_i$ and $\ueq=\ueq_i$ for $i=1$. For a comparison of step motion by this formula with neglect of the $\Ld\kappa$ term to the boundary integral formulation of Eq.~\eqref{eqs:ies_for_omegai_bc}, the reader is referred to Sec.~\ref{subsec:validity}. 

A few remarks on Eqs.~\eqref{eq:main_flux_result} and~\eqref{eq:main_flux_1step} are in order. First, the local step curvature enters these approximations in the following two distinct ways: (i) thermodynamically, by the equilibrium step edge (shifted) concentration $\ueq_j$ ($j=i,\,i\pm 1$) which depends on the step stiffness through the step chemical potential; and (ii) kinetically, through the terms of the form $\Ld \kappa_j$ and $(1 \mp \bwt  \kappa_j)^{-1/2}$. We emphasize that there is no approximation in our use of $\ueq_j$ with the exception of its slow variation along the step edge. A noteworthy feature of our analysis is that it singles out the above kinetic contributions naturally, by relating them to the asymptotic regime of strong desorption. Second, as is expected by mere inspection, the scaled step velocity $v_{i,\perp}/\ueq_i$ of an isolated step in Eq.~\eqref{eq:main_flux_1step} is invariant under the interchange of $\Lad^{\pm}$ provided the sign of curvature $\kappa_i$ is reversed. 

Third, it is tempting to compare the results of this section to Eq.~\eqref{eq:radial_flux_asym} of the radial case. We realize that the step velocity $v_{i,\perp}(\mx)$ here reduces to the one obtained for the radial geometry if the (non-dimensional) quantity $\Ld \kappa_j$ in Eq.~\eqref{eq:main_flux_result} is neglected while $\Lad^{\pm}$ are kept fixed. In this vein, we obtain  
\begin{widetext}
\begin{align} \label{eq:main_flux_result-leading}
	v_{i,\perp} &\simeq 
	D_s A \left \{ -\ueq_i \left (\frac{1}{\Lad^+ + \Ld} 
	+ \frac{1}{\Lad^- + \Ld} \right ) + \ueq_{i+1}\,  (1-\bwt_{i} \kappa_{i+1})^{-1/2}  
	\frac{2\Ld}{\left (\Lad^+ + \Ld \right) \left (\Lad^- + \Ld\right )}\  e^{-\frac{\bwt_{i}}{\Ld}} \right. \nonumber \\
	& \qquad \left. + \ueq_{i-1}\, (1+\bwt_{i-1} \kappa_{i-1})^{-1/2}   
	\frac{2\Ld}{\left ( \Lad^+ + \Ld \right ) \left(\Lad^- + \Ld \right)} e^{-\frac{\bwt_{i-1}}{\Ld}}\right \}~,
\end{align}
\end{widetext}
\end{subequations}
which forms a generalization of the step velocity law of the radial setting (Sec.~\ref{subsec:explicit-radial}).
This formula reduces to Eq.~\eqref{eq:radial_flux_asym} if the steps are concentric circles with radii $r_i=1/\kappa_i=\Ld R_i$ in our notation, where $\bwt_i=r_{i+1}-r_i=\Ld (R_{i+1}-R_i)$; see Sec.~\ref{subsec:explicit-radial}. Note in passing that the aforementioned discrepancy between the formulas of the two geometries, for nonzero yet small $\Ld\kappa_j$, manifests only in the respective correction terms. This discrepancy can be remedied if we include more terms in the expansions used for the Bessel functions in the radial case.

We stress that our analysis treats each step curve as a given smooth boundary, and formally produces an asymptotic formula for the step velocity for small enough diffusive length $\Ld$. Some aspects of our results, particularly the relevant kinetic lengths, are discussed in Sec.~\ref{subsec:kin-lengths}.  The consistency of this approach with the well-posedness of step motion, when each step edge is viewed as a free boundary, is not addressed by our approach. The kinetic role of the step curvature, as this appears in our asymptotic results, is further discussed in Sec.~\ref{subsec:limitations}. \color{black}

\section{Additional kinetic effects} 
\label{sec:extensions}
In this section, we outline extensions of our formalism. In particular, we incorporate the kinetic effects of step edge diffusion and step transparency (permeability) into the boundary integral equations. We
show how these modifications affect the step velocity law in the limit of strong desorption.

\subsection{Step edge diffusion}
\label{subsec:edge-diff}
In step edge diffusion, atoms that have already attached to the step may move along its edge with a possibly orientation-dependent diffusivity, $D_e$~\cite{OPL2001,Dankeretal2004,Paulinetal2001,Krug_in_Voigt}.  
This process can be included in the mass conservation statement for the motion of steps. Accordingly, the step velocity in the direction of the local normal vector $\nv$ pointing to the lower terrace becomes
\begin{equation}\label{eq:st-edge_diff}
v_\perp=A J_\perp+\frac{a}{\Ld^2c_s}\,\partial_\varsigma\left(D_e \partial_\varsigma\ueq\right)~.
\end{equation}
Here, $J_\perp=D_s \nv\cdot \{(\nabla C)^+-(\nabla C)^-\}$ is the total normal flux into the step, $\partial_\varsigma$ is the dimensionless tangential derivative along the step edge (where the arclength is scaled by $\Ld$), $c_s$ is the equilibrium adatom density of a straight step, and $A$ is the atomic area. Evidently, for a given step curve, the normal flux $J_\perp$ can be determined from the boundary value problem for adatom diffusion on the adjacent terraces (see Sec.~\ref{subsec:diffusion}). 
This problem for $J_\perp$ can be tackled separately from the edge diffusion process. Of course, in the course of time evolution, edge diffusion (via $D_e$) alters the step shape.  

For our purposes, the velocity $v_\perp$ has two distinct contributions. In the limit of strong desorption, the term pertaining to $AJ_\perp$ comes from the framework of Sec.~\ref{sec:main}, and is given by Eq~\eqref{eq:main_flux_result}. On the other hand, the edge diffusion term of Eq.~\eqref{eq:st-edge_diff} involves tangential derivatives of $\ueq$. If the step chemical potential $\mu$ is dominated by the step stiffness $\tilde\gamma$ with $|\mu|\ll T$, we can formally approximate $\partial_\varsigma\ueq \simeq c_s T^{-1} \partial_\varsigma (\tilde\gamma \kappa)$ in Eq.~\eqref{eq:st-edge_diff}. Hence,  the emerging velocity $v_\perp$ depends on the local curvature $\kappa$ and the derivative $\partial_\varsigma(D_e\partial_\varsigma(\tilde\gamma\kappa))$.

  Intuitively, we expect that step edge diffusion has a regularizing effect on the adatom equilibrium density, $\ceq$. In particular, 
say, for constant $D_e$, this process causes $\mu$ and thus $\ceq$ to vary more slowly along the step. This effect should improve the accuracy of our asymptotic formulas. 
\color{black}

\subsection{Step transparency}
\label{subsec:transp}
\begin{figure}
\begin{center}
\includegraphics[scale=0.35,trim=0.72in 0.5in 0in -0.2in]{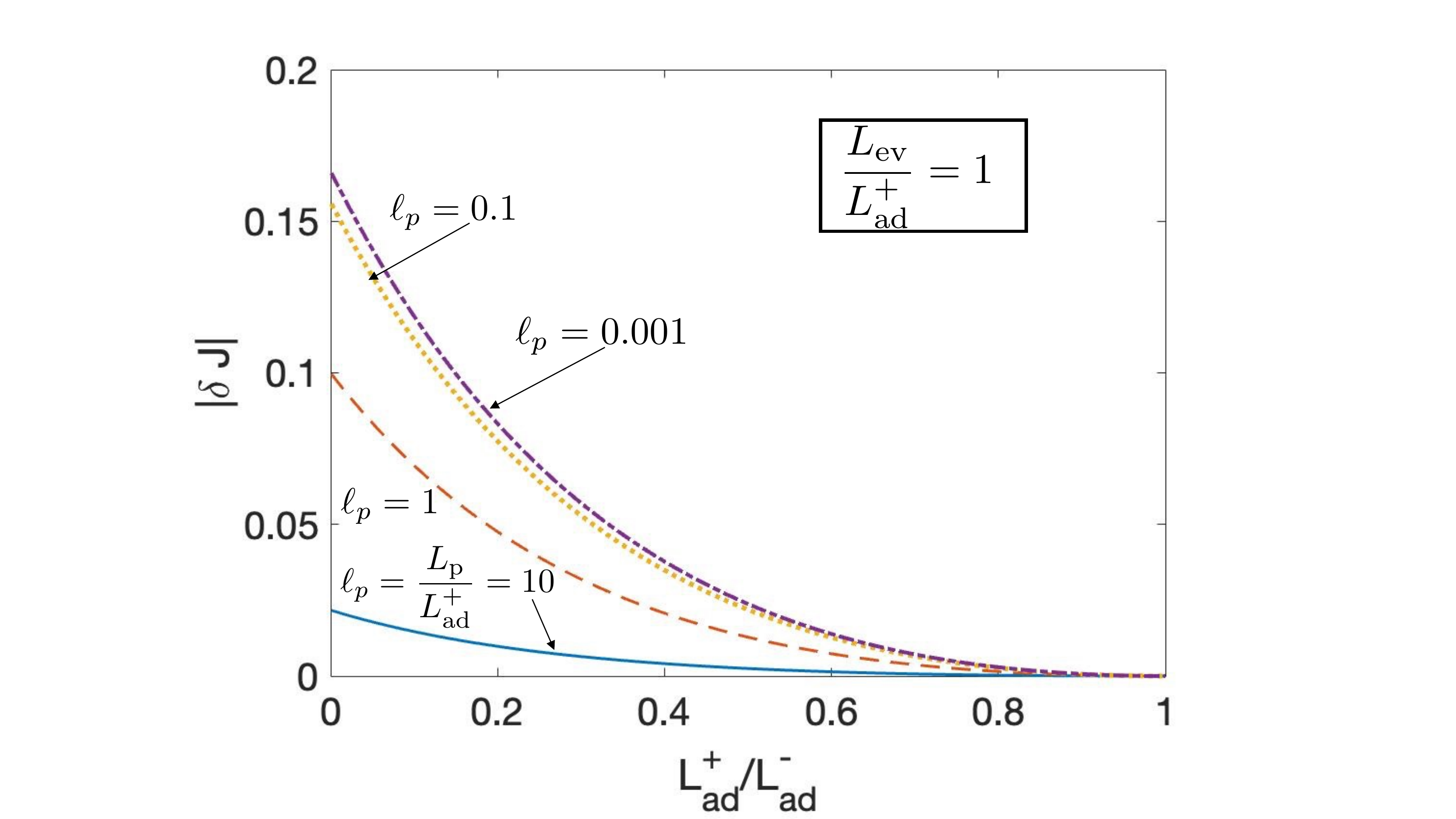}
\end{center}
\caption{Plots of magnitude of normalized adatom flux $\delta J$ defined by Eq.~\eqref{eq:deltaJ} as a function of Ehrlich-Schwoebel barrier parameter $\Lad^+/\Lad^-$. We use distinct values of step permeability parameter $\ell_p=\Lp/\Lad^+$ ($\ell_p=10,\,1,\,0.1,\,0.001$); and desorption parameter $\Ld/\Lad^+=1$. Smaller values of $\ell_p$ imply more permeable steps.}
\label{fig:deltaJ-comp}
\end{figure}

So far, we have assumed that adatoms are exchanged between neighboring terraces only via atom attachment and detachment at steps; see Eq.~\eqref{eq:bcf_bc_general} with $\Lp=\infty$. In the presence of step permeability, when the length $\Lp$ is finite, we allow for the direct mass exchange between terraces~\cite{Zangwill1992}. In principle, this mechanism eventually alters the total adatom flux into the step, and thus the step velocity. Intuitively, we expect this effect to be pronounced when strong enough step permeability (small enough length $\Lp$) is combined with sufficiently high step edge barrier asymmetry. This behavior is simply demonstrated in Fig.~\ref{fig:deltaJ-comp}. \color{black} Here, we capture this effect in a prototypical setting by considering only the term that couples the step velocity with the equilibrium adatom density of the {\em same} step. 

The procedure used in Sec.~\ref{subsec:asymptotics} for the setting with impermeable steps can be applied when $\Lp$ is finite. We will explicitly show how the length $\Lp$ affects the emerging step velocity law for strong desorption by neglecting kinetic step-step interactions in the formalism.

Consider the geometry of Fig.~\ref{fig:Green-geom}. By boundary condition~\eqref{eq:bcf_bc_general} we obtain 
\begin{subequations} \label{eqs:perm_u}
\begin{align}
\tilde C^+ & = \frac{\Lad^+ \Lad^- \left [ \left(\frac{\partial C}{\partial \nu}\right)^+ - \left(\frac{\partial C}{\partial \nu}\right)^- \right ] + \Lad^+ \Lp \left(\frac{\partial C}{\partial \nu}\right)^+}{\Lad^+ + \Lad^- + \Lp}~, \label{eq:perm_u+} \\
\tilde C^- & = \frac{\Lad^+ \Lad^- \left[ \left(\frac{\partial C}{\partial \nu}\right)^+ - \left(\frac{\partial C}{\partial \nu}\right)^- \right] - \Lad^- \Lp \left(\frac{\partial C}{\partial \nu}\right)^-}{\Lad^+ + \Lad^- + \Lp}~, \label{eq:perm_u-}
\end{align}
\end{subequations}
where $\tilde C=C-\ueq$ denotes the deviation of the shifted adatom density $C$ from the equilibrium value $\ueq$ at point $\mx$ of a step curve. Notice that each of the limiting values $\tilde C^\pm$ depends on the fluxes at both sides of the same step. Hence, the procedure of Sec.~\ref{subsec:asymptotics} implies that the resulting system of boundary integral equations for the one-sided normal fluxes couples all steps simultaneously, rather than merely coupling the step to its nearest neighbors. In other words, in the presence of step permeability Eq.~\eqref{eqs:ies_fin_approx} should be replaced by a system of equations that couples all steps. Since our focus here is on local approximations for strong desorption, we omit writing out the resulting system.

In the aforementioned framework of approximations, we neglect all terms that produce couplings of the fluxes at a given step to those of adjacent steps.  Thus, the system of equations for permeable steps becomes local, similar to the situation described by Eq.~\eqref{eq:main_flux_1step}. \color{black} 

We further simplify the governing equations by neglecting the kinetic curvature contributions to the adatom fluxes. Thus, our asymptotics yield 
\begin{equation*} 
C^\pm\simeq \mp \Ld \left(\frac{\partial C}{\partial \nu}\right)^{\pm}~,
\end{equation*}
at point $\mx$ of a step curve. The combination of the last relation and Eq.~\eqref{eqs:perm_u} furnishes the one-sided fluxes (cf. Fig.~\ref{fig:Green-geom})
\begin{subequations} \label{eqs:permeable_fluxes}
\begin{align}
-D_s\left(\frac{\partial C}{\partial \nu}\right)^+ & \simeq \frac{D_s}{(1-\alpha^+) \Lad^+ + \alpha^+ \Lad^- + \Ld} \,\ueq, \label{eq:permeable_fluxes-a}\\
-D_s\left(\frac{\partial C}{\partial \nu}\right)^- & \simeq  -\frac{D_s}{(1-\alpha^-) \Lad^- + \alpha^- \Lad^+ + \Ld}\, \ueq, \label{eq:permeable_fluxes-b}
\end{align}
where the (non-dimensional) kinetic parameters $\alpha^\pm$ are 
\begin{equation}\label{eq:alpha-def}
\alpha^\pm=\left(1+\frac{\Lad^\mp}{\Lad^\pm} + \frac{\Lp}{\Lad^\pm} + \frac{\Lp}{\Ld}\frac{\Lad^\mp}{\Lad^\pm}\right)^{-1}~.
\end{equation}
\end{subequations}
Note that $0<\alpha^\pm <1$, and $\alpha^\pm\to 0$ if $\Lp\to \infty$.

We comment on Eq.~\eqref{eqs:permeable_fluxes} for fixed $\ueq$. Evidently,
the flux in each side of the permeable step depends on an effective attachment-detachment length equal to $(1-\alpha^\pm)\Lad^\pm+\alpha^\pm \Lad^\mp$, which is a convex-type combination of the original lengths $\Lad^+$ and $\Lad^-$. 
Our asymptotic formulas reveal that if $\Lad^- > \Lad^+$, which occurs for a positive Ehrlich-Schwoebel barrier~\cite{Ehrlich,Schwoebel}, the magnitude of the normal flux from the upper terrace (`$-$' side of step) increases as $\Lp$ decreases, i.e., as steps become more transparent. In contrast, the normal flux into the lower terrace (`$+$' side of step) decreases. 
\color{black}

Next, we describe the difference of the total adatom flux into a permeable step from its counterpart for an impermeable step. We choose to express this difference in units of $J^0=D_s \ueq/\Lad^+$. Therefore, we consider the quantity
\begin{equation}\label{eq:deltaJ}
	\delta J= \frac{J_\perp^{\text{per}}-J_\perp^{\text{imp}}}{J^0}~,
\end{equation}
where $J_\perp^{\text{K}}=D_s \{(\partial_\nu C)^+-(\partial_\nu C)^-\}$ for permeable (K=`per') or impermeable (K=`imp') steps. In this computation, we invoke approximate formulas~\eqref{eq:permeable_fluxes-a} and~\eqref{eq:permeable_fluxes-b}. In Fig.~\ref{fig:deltaJ-comp}, we display plots of $|\delta J|$ as a function of the parameter $\Lad^+/\Lad^-$ which signifies the positive Ehrlich-Schwoebel barrier, for different values of $\ell_p=\Lp/\Lad^+$, which measures step transparency, with fixed value of $\Ld/\Lad^+$ ($\Ld/\Lad^+=1$) which expresses the desorption strength. Note that small values of $\Lad^+/\Lad^-$ amount to high step edge edge barrier asymmetry. Small values of $\ell_p$ with fixed length $\Lad^+$ imply strong step transparency. \color{black} We conclude that the effect of step permeability on the (scaled) total flux into the step is favored by small values of $\Lad^+/\Lad^-$, when the barrier asymmetry is appreciable. In addition, the ratio $\Ld/\Lad^+$ must not be too small. 
These results can be refined analytically by a close inspection of our approximate formulas for the flux. We choose not to pursue this task here.

\section{Discussion}
\label{sec:numerics}
In this section, we discuss the validity and implications of our approach.
By use of an ad hoc anisotropic step free energy, we check numerically the accuracy of our asymptotic formula for the velocity of a single step  (see Sec.~\ref{subsec:validity}). We also discuss implications of our main results, particularly the significance of the kinetic lengths entering the simplified, local step velocity law (Sec.~\ref{subsec:kin-lengths}). Furthermore, we outline limitations of our approach, which inspire other, open problems (Sec.~\ref{subsec:limitations}). For example, we discuss the possible breakdown of our hypothesis for smooth step curves; the character of possible correction terms, to higher orders in the length $\Ld$; and the effect of surface electromigration which can modify the integral equation formalism.  

\subsection{On the numerical validation of our approximations}
\label{subsec:validity}
Next, we carry out numerical simulations in order to check the accuracy of asymptotic formula~\eqref{eq:main_flux_1step} for an isolated step (if $N=1$) by neglecting the term $\Ld \kappa$. \color{black} In particular, using suitable coordinates $(r,\vartheta)$ for the step curve, we compare the approximate prediction for $r(\vartheta)$ at sufficiently long times to the corresponding result computed from solving Eq.~\eqref{eqs:ies_for_omegai_bc} of the boundary integral formalism by quadrature. Our numerics capture the late-time  morphological evolution of the step curve, after any transient effects become negligible. \color{black}

Let us review briefly the notion of the anisotropic step free energy~\cite{JeongWilliams1999,Krug_in_Voigt}. For an isolated step $\Gamma$, the total energy is
\begin{equation*}
E_{\rm{st}}=\int_\Gamma \gamma(\boldsymbol \nu(\my))\, ds_\my~,
\end{equation*}
where $\gamma(\boldsymbol\nu)$ is the step free energy per unit length (line tension) at the point $\my$ through the local normal vector $\boldsymbol\nu$ to $\Gamma$. Assuming that curve $\Gamma$ is described locally as the graph of function $x(y)$, we write $\mu=A \frac{\delta E_{\rm{st}}}{\delta x}=A \tilde{\gamma} \kappa$, where $\tilde\gamma$ is the step stiffness. Abusing notation, we write $\tilde{\gamma}(\vartheta)=\gamma(\vartheta)+\gamma''(\vartheta)$ where $\vartheta$ is the angle between the normal vector to $\Gamma$ and the positive $x$ axis; the prime denotes differentiation with respect to the argument. Recall that $\mu$ enters $\ueq$ via relation~\eqref{eq:ceq-BCF} with $\ceq=\ueq + F\tau$; here, we have $\ceq=\ceq_1$ (since $N=1$).

Motivated by~\cite{LeeThorp}, in numerics we use the following model of step stiffness as a function of the step orientation angle $\vartheta$:
\begin{equation}\label{eq:stiff-choice}
\frac{A \tilde{\gamma}(\vartheta)}{T}=\{1-0.99\cos(6\vartheta)\}L
\end{equation}
where $L$ has the dimension of length.
We use small enough yet nonzero value of $F\tau$ so that adatom diffusion reaches the quasi-steady regime at long times. \color{black} 
If we start from a smooth initial step shape close to a circle, we expect that the step curve approaches a limit, which we view as a `kinetic Wulff shape' of the growth process~\cite{Sekerka2005}. By our choice of $\tilde\gamma(\vartheta)$ the kinetic Wullf shape will resemble a regularized hexagon, with slightly rounded corners and slightly curved edges. 

We proceed to elaborate on our numerical simulations. We use an initial circular step shape with radius $R^0=L$; cf. Eq.~\eqref{eq:stiff-choice}.  Subsequently, we scale all length scales and spatial coordinates of our system by this $R^0$; alternatively, set $R^0=1=L$ throughout. 
We assume that the step edge barrier is symmetric, and thus set $\Lad^-=\Lad^+=\Lad$ with $\Lad/R^0=1$. We also take $(R^0)^2F\tau=3$ and $c_s=(R^0)^{-2}$. \color{black}
Let us scale time by $t_0$ where $D_s t_0/[R^0(\Lad+\Ld)]=0.5$. 

\begin{figure}
\begin{center}
\includegraphics[scale=0.115,trim=1in 0.5in 0in -0.2in]{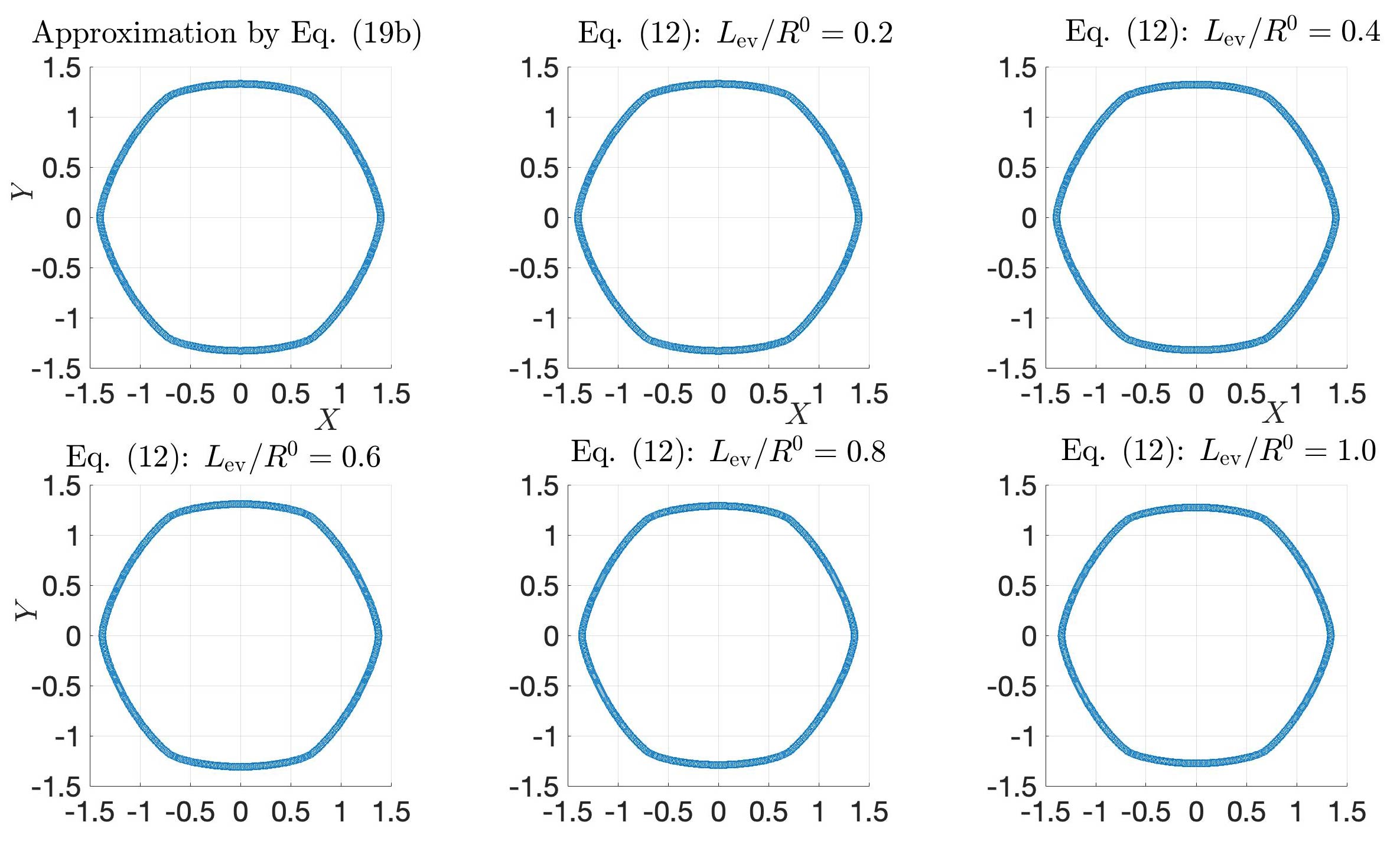}
\caption{Snapshots of step shape at long time via numerical simulations by use of: asymptotic formula~\eqref{eq:main_flux_1step} with $\Ld\kappa=0$ (top left panel); and boundary integral equations~\eqref{eq:ies_for_omegai_bc-1} and~\eqref{eq:ies_for_omegai_bc-2} via quadrature. A single isolated step is used. The same scaled time $t/t_0$ is used for the snapshots. \color{black} The initial shape is a circle of radius $R^0$. The axes correspond to scaled spatial coordinates $(x,y)$: $X=x/R^0$ and $Y=y/R^0$.
Top left panel: The result is independent of parameter $\Ld/R^0$. Remaining plots (boundary integrals by quadrature): $\Ld/R^0=0.2,\,0.4,\,0.6,\,0.8,\,1.$}\label{fig:snapshots}
\end{center}
\end{figure}

With our choice of parameters, our goal is to indicate that desorption may plausibly enable the emergence of local geometric laws during step motion, under suitable conditions. The physical roles of these parameters can be outlined as follows. 
The stiffness $\tilde\gamma$ is modeled phenomenologically to capture the  formation of smoothed corners in the step shape. We allow for barely enough deposition flux $F$ on the surface from above so that growth can balance out desorption at intermediate times, although growth becomes appreciable at long times. We apply mixed kinetics in the sense that surface diffusion in the inner terraces is balanced out by the attachment and detachment of atoms at steps. In this regime, we show numerically that there exists a time window in which local geometric laws can occur during step shape evolution. In practice, this time interval should be controlled by $t_0$, which in principle depends on the temperature and material, and the initial geometry. We have not made any effort to implement parameters of specific materials here.
\color{black}

In our numerical simulations, we let the step shape evolve with time according to approximate Eq.~\eqref{eq:main_flux_1step}; or, alternatively, according to the boundary integrals of Eq.~\eqref{eqs:ies_for_omegai_bc} for a few distinct values of the parameter $\Ld/R^0$. The respective step shape, at the same scaled time $t/t_0$, \color{black} is depicted in Fig.~\ref{fig:snapshots}. We observe that in each case the step shape has practically converged to a steady state. \color{black} There are no significant differences between any of the generated shapes for the chosen values of $\Ld/R^0$ (see comment below). Note that our leading-order formula~\eqref{eq:main_flux_1step} for the step velocity, with the kinetic contribution $\Ld\kappa$ set to zero, produces the same late-time shape, regardless of the value that we use for $\Ld/R^0$. This result is shown in the top left panel of Fig.~\ref{fig:snapshots}. The independence of this outcome from $\Ld/R^0$ is expected since the simplified step velocity from Eq.~\eqref{eq:main_flux_1step} depends on the parameters $D_s$, $\Lad$ and $\Ld$ alone, through the ratio $D_s/(\Lad+\Ld)$ which is held fixed in our numerics.  \color{black}

A few more comments on Fig.~\ref{fig:snapshots} are in order. An inspection of the plots based on Eq.~\eqref{eqs:ies_for_omegai_bc} indicates that larger values of $\Ld/R^0$ cause slightly less growth, as expected intuitively. Overall, the comparison of asymptotic and boundary integral equation predictions is surprisingly favorable even for values of $\Ld/R^0$ close to unity, as we discuss below. We should point out that the underlying error is expected to increase significantly for large enough values of  $\Ld/R^0$. Presumably, our leading-order asymptotic formula for the step velocity breaks down if $\Ld/R^0$ becomes sufficiently large. \color{black}

\begin{figure}
\begin{center}
\includegraphics[scale=0.133,trim=1.7in 0.5in 0in -0.2in]{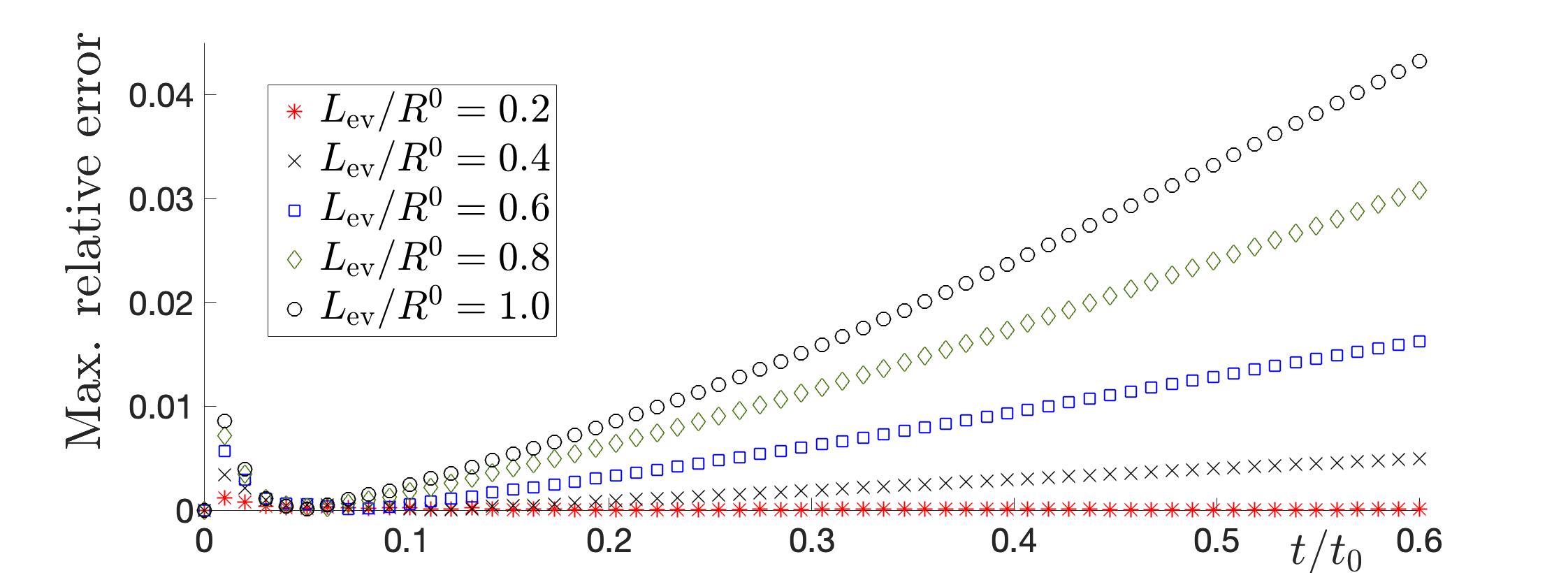}
\caption{Plots of maximum relative error over scaled time in regard to  asymptotic formula~\eqref{eq:main_flux_1step} with $\Ld\kappa=0$. Numerical simulations based on Eq.~\eqref{eq:main_flux_1step} are compared to simulations based on quadrature for the boundary integral formalism of Eq.~\eqref{eqs:ies_for_omegai_bc}. The following values of $\Ld/R^0$ are used: $\Ld/R^0=0.2$ (asterisk), 0.4 (cross), 0.6 (square), 0.8 (diamond), and 1 (circle).}\label{fig:errors}
\end{center}
\end{figure}

We now briefly describe the error from the use of asymptotic formula~\eqref{eq:main_flux_1step} with $\Ld\kappa$ set to zero, for distinct values of $\Ld/R^0$. For this purpose, we numerically compute the maximum relative error in $r(\vartheta)$ versus scaled time $t/t_0$, by comparison to the corresponding result of the boundary integral formalism according to Eq.~\eqref{eqs:ies_for_omegai_bc} via quadrature. This relative error is shown in Fig.~\ref{fig:errors}, for several values of $\Ld/R^0$. Our results confirm that the relative error increases with $\Ld/R^0$. A surprising aspect of this comparison is that even for $\Ld/R^0=1$ the error does not exceed about $5\%$. For our chosen parameter values and geometry, the step growth is relatively rapid which in turn favors small relative error. On the other hand, a choice of parameters that causes relatively slow growth would result in larger relative error.

The plot in Fig.~\ref{fig:errors} also indicates that the maximum relative error reaches a minimum at $t/t_0\simeq 0.05$. We have not been able to provide a quantitative explanation for this lack of monotonicity of the relative error with time. The minimum appears to be a small effect overall, but tends to be a little more pronounced for larger values of $\Ld/R^0$.  We stress that if $\Ld/R^0$ is small, approximately equal to 0.2 or smaller, the computed maximum relative error is negligible, and its non-monotonicity with time is barely evident in our numerics. This is the asymptotic regime that allows for the emergence of geometric motion laws.

\subsection{Prediction: Effective kinetic lengths}
\label{subsec:kin-lengths}

In this subsection, we review some of our results and discuss their possible implications. We place emphasis on the emergence of kinetic lengths in the simplified velocity law of a single isolated step under strong desorption. 

In the absence of step permeability, consider Eq.~\eqref{eq:main_flux_1step} for a single, isolated step. Now set the kinetic contribution $\Ld\kappa$ equal to zero. The step velocity has the form
\begin{subequations}\label{eqs:one-step-eff-law}
\begin{equation}\label{eq:one-step-eff-law-1}
 v_\perp \simeq -AD_s \Lef^{-1} \ueq~, 	
\end{equation}
where $\Lef$ is an effective kinetic length defined as
\begin{equation}\label{eq:one-step-eff-law-2}
 \Lef=\left(\frac{1}{\Lad^++\Ld}+\frac{1}{\Lad^-+\Ld}\right)^{-1}~.	
\end{equation}
\end{subequations}
In the above, the lengths $\Lad^{\pm}=D_s/k^\pm$ express the step edge barrier asymmetry (for $k^+\neq k^-$). Notice that $\Lef$ is a harmonic-type mean of the diffusion lengths $\Lad^\pm+\Ld$. Each of these two lengths is the average distance that an adatom has to travel via hopping on the respective terrace adjacent to the step edge.
We stress that Eq.~\eqref{eqs:one-step-eff-law} reduces to the classic result of BCF for a circular step~\cite{BCF} without an Ehrlich-Schwoebel barrier and if $\ueq$ becomes a linear function of the local step curvature, $\kappa$ (see Sec.~\ref{subsec:explicit-radial}). 

Notably, our analysis provides an extension of Eq.~\eqref{eqs:one-step-eff-law} to the setting with step permeability; cf. Eq.~\eqref{eqs:permeable_fluxes}. The idea suggested by this extension is that the step velocity remains of the same form, Eq.~\eqref{eq:one-step-eff-law-1}, yet with an effective length $\Lef$ that introduces a `renormalization' of the step edge asymmetry lengths, $\Lad^\pm$. Our results show the following related substitutions: 
\begin{equation}
 \Lad^\pm \longrightarrow (1-\alpha^\pm) \Lad^\pm +\alpha^\pm \Lad^{\mp}~,	
\end{equation}
where the constants $\alpha^\pm$ are defined in Eq.~\eqref{eq:alpha-def}. Recall that a key assumption in our analysis, which allows for this correspondence, is that kinetic interactions between steps are negligible because of the effect of strong desorption across wide enough terraces. As a result, the intrinsically nonlocal effect of step transparency becomes effectively local. Recall Fig.~\ref{fig:deltaJ-comp}, in which the settings with permeable and impermeable steps are compared via the scaled normal flux $\delta J$.

\subsection{Limitations}
\label{subsec:limitations}
Next, we outline limitations of our approach. In particular, we discuss the assumption that the step curve is fixed and smooth. We also remark on the character of the correction terms, which would result in modifications of our leading-order asymptotic formula for the step velocity. Furthermore, we provide the example of an additional kinetic effect, namely, surface electromigration, which would modify a part of the integral equation formalism.

\subsubsection{Smoothness of step curve}
\label{sssec:smooth}

A major limitation of our formalism is the underlying assumption about the geometry: the step is supposed to be represented by a (given) smooth curve~\cite{Note-facets}. In fact, to be more precise, each step curve must be twice continuously differentiable with respect to the arclength on the crystal reference plane. This hypothesis is questionable in many situations. For instance, on a crystal surface with strongly anisotropic step line tension, the step geometry can become singular, e.g., have facets (straight lines). 

We expect that our formalism can still be applied along such facets on steps with the appropriate modification of the local curvature~\cite{LeeThorp}; but would presumably break down near corners between facets. \color{black} This pathology is partly caused by the poor resolution of the distance between two points on the opposite sides of the corner, since this distance is not differentiable with respect to the step arclength at the corner. Consequently, the approximation 
$|\my_j(\varsigma)-\my_j(\varsigma_0)| \simeq |\varsigma-\varsigma_0|\Ld$ invoked in Sec.~\ref{sssec:sd-potentials}, where now the value $\varsigma=\varsigma_0$ corresponds to the corner position, becomes inaccurate when $|\varsigma-\varsigma_0|$ is of the order of unity or smaller. Furthermore, if the step line tension $\gamma$ has an explicit dependence on the step orientation then $\ueq$ can be discontinuous at the corner between two facets. 
Thus, the replacement of $\ueq$ by a constant, which we applied in our leading-order approximation scheme (Sec.~\ref{subsec:asymptotics}), is expected to fail near the corner.

A plausible remedy would be to split the integration for the relevant single- and double-layer potentials at the position of the corner. We also need to adopt a more sophisticated approximation for $\ueq$ along the step. The simplest possible scenario would be to  replace $\ueq$ by a different constant on each side of the corner. This problem is not studied here.\color{black}

If the step line tension $\gamma(\vartheta)$, as a function of the orientation angle $\vartheta$, has corner singularities at each of its local minima and the step edge is initially faceted at the corresponding orientations, the above approximation about $\ueq$ being piecewise constant becomes an exact property. This situation is encountered in~\cite{LeeThorp} where the authors invoke the notion of the weighted mean curvature. Their
computation relies on the properties that the facet is perfectly flat and admissible perturbations of the step shape preserve the character of this facet.  Our analysis cannot address this setting.
\subsubsection{Correction terms}
\label{sssec:corrections}
Our analysis so far mostly concerns the derivation of leading-order formulas for step velocities. We  have also indicated the kinetic effect of curvature through the term $\Ld\kappa$ as well as the kinetic interactions of a step with its nearest neighbors via effective terrace widths; see Secs.~\ref{subsec:emergence} and~\ref{sec:extensions}. This treatment points to at least two questions. One question is: How can one derive higher-order terms of the asymptotic expansion for the step velocity, with given smooth step shape? Another, more challenging question is: Can such higher-order terms be used reliably to describe the step morphological evolution? We briefly discuss these issues.

Regarding the first issue, for a smooth step shape, one can in principle derive correction terms to arbitrary order in the length $\Ld$ via a suitable change of variable in the integrals for the requisite potentials. We outline the procedure here for the interested reader. The core idea relies on a standard but elaborate procedure of classical asymptotics~\cite{Erdelyi-book}. To convey this idea, let us restrict attention to the single-layer potential term $\mts_i[f](\mx)$ when the point $\mx$ lies in step $\Gamma^{(i)}$; $f$ is the normal flux. Suppose that the step curve is parametrized by the scaled (signed) arclength $\varsigma$. For sufficiently small $\Ld$, we need to expand part of the integrand around some value $\varsigma=\varsigma_0$, e.g., $\varsigma_0=0$. 
For the derivation of higher-order terms, the approximation $|\my_i(\varsigma)-\my_i(0)| \simeq |\varsigma|\Ld$, which we applied previously for the argument of the kernel (Sec.~\ref{sssec:sd-potentials}), is no longer adequate. Instead, we can handle this case by changing the integration variable from $\varsigma$ to $\mathcal R$ according to 
$ \mathcal R(\varsigma)=|\my_i(\varsigma)-\my_i(0)|/\Ld$. This choice requires splitting the starting integral in way that renders the distance function  $\mathcal R(\varsigma)$ one-to-one in each domain of the $\mathcal R$-integration. 

Subsequently, each integral can produce an expansion in powers of $\Ld$ as follows. We can write the part of each integrand other than the kernel as a polynomial in $\mathcal R$, and then integrate term by term. This task is carried out via the approximation of the function $f(\varsigma)$ as a polynomial in $\varsigma$ via a Taylor expansion. By inversion of $\mathcal R(\varsigma)$, we can determine the respective polynomial in $\mathcal R$ for $f$~\cite{Erdelyi-book}.  This procedure suggests that, in the case of a single step, our analysis is reasonable provided that the flux normal to the step does not vary appreciably over arclengths of the order of $\Ld$. The details of this procedure are omitted here.

Despite the systematic derivation of higher-order terms in $\Ld$ for the step velocity, as indicated above, their role in the step morphological evolution is not addressed. For instance, consider Eq.~\eqref{eq:main_flux_1step} by regarding the kinetic contribution $\Ld\kappa$ as the first correction. A natural question is whether the motion law for the step in the presence of this correction is well posed or not. This aspect of our asymptotic results, namely, the implications of the asymptotic expansion in $\Ld$ for the actual step dynamics, where the geometry evolves with time and thus forms part of the overall solution, is left unresolved.

\subsubsection{Another kinetic effect: Surface electromigration}
\label{sssec:electrom}
Next, we discuss the kinetic effect of surface electromigration which requires a modification of our formalism.
In the presence of an external electric field $\boldsymbol E$, the positively charged adatoms are forced to move in its direction. This motion causes a drift velocity $\mathbf{v}_E$ which is given by~\cite{Fu1997,Dufayetal2007,QuahDM2010} 
\begin{equation*}
	\mathbf{v}_E=\frac{D_s(Z^*e) \boldsymbol E}{T}~.
\end{equation*}
Here, the constant $Z^* e$ is the effective adatom charge; $|Z^*|$ is greater than unity for metals but can be quite small for semiconductors~\cite{Dufayetal2007}. For simplicity, let us assume that $\boldsymbol E$ is constant; thus, $\mathbf{v}_E$ is constant. The electric field produces a convective term in the diffusion equation for adatoms via the drift velocity. Hence, the shifted adatom concentration $C$ on terraces satisfies the equation
\begin{equation*}
\Delta C = \Ld^{-2} C + D_s^{-1}\mathbf{v}_E \cdot \nabla C~. 
\end{equation*}
In addition, we impose Robin-type boundary conditions~\eqref{eq:bcf_bc_general} for attachment and detachment of atoms at the step edges.

If we invoke the Green function $G(\mx, \my)$ (Sec.~\ref{subsec:formalism}), the standard procedure of the integral equation formalism yields
\begin{align}
-C(\mx) & = \frac{\mathbf{v}_E}{D_s} \cdot\int_{\Omega^{(i)}} G(\mx,\my)\,  \nabla_\my C(\my) \: d\my \nonumber \\
& + \mts_{i+1}\biggl[\biggl(\frac{\partial C}{\partial \nu}\biggr)^- \biggl](\mx) - \mts_i\biggl[\biggl(\frac{\partial C}{\partial \nu}\biggr)^+\biggr](\mx)  \nonumber \\
& - \left \{ \mtd_{i+1}[C^-](\mx) - \mtd_i[C^+](\mx) \right \}~,\quad \mx\ \mbox{in}\ \Omega^{(i)}~; \label{eq:electro_ie}
\end{align}
cf. Eq.~\eqref{eq:Green-form} in which we omit the $\pm$ superscripts.
We can now apply integration by parts to remove the gradient operator $\nabla_{\my}$ from $C$ in the integral of the first line in Eq.~\eqref{eq:electro_ie}. Note that the singularity of the ensuing term $\nabla_\my G(\mx,\my)$, when $\mx=\my$, does not manifest since $\mx$ is not let to coincide with $\my$ as yet. We can then pull the gradient operator  out of the respective integral according to
\begin{equation*}
- \int_{\Omega^{(i)}} \nabla_\my G(\mx,\my) C(\my)\,d\my = \nabla_\mx \left [ \int_{\Omega^{(i)}} G(\mx,\my)  C(\my) d \my \right ].
\end{equation*}
This equation may not be further simplified to yield boundary integrals along steps. Hence, the use of the same Green function $G$ via Eq.~\eqref{eq:electro_ie} requires computing the shifted concentration field $C$ on the whole terrace, in contrast to the spirit of boundary integral equations in this paper. 

Nonetheless, a boundary integral formalism can be derived for this case by use of a different Green's function $\check G(\mx, \my)$, which accounts for the drift velocity $\mathbf{v}_E$. This $\check G$ should obey the equation $\big\{\nabla_\mx\cdot\big(\varphi(\mx)\nabla_\mx\big)-\Ld^{-2}\big\}\check G(\mx,\my)=\delta(\mx-\my)$, where 
$\varphi(\mx)=\exp(-D_s^{-1}\mathbf{v}_E\cdot \mx)$ is an integrating factor implied by the diffusion equation with a drift for density $C$. This definition leads to a boundary integral formalism under electromigration analogous to that of Sec.~\ref{subsec:formalism}. The analysis of this problem will be the subject of future work.

\section{Conclusion} 
\label{sec:conclusion}

In this paper, we derived simplified formulas for the velocities of line defects (steps) with fixed shapes on a crystal surface under growth conditions below the roughening transition. The starting point is the BCF model in the quasi-steady regime, enriched with kinetic conditions for atom attachment and detachment at steps as well as step permeability and diffusion along steps. Our main assumptions are that the diffusion length $\Ld$ due to evaporation on terraces is small compared to the step linear size and radius of curvature, and each step curve is smooth. 
We recognize that a  narrow boundary layer of adatom diffusion develops near the step edges. Hence, the intrinsically nonlocal mechanism of adatom diffusion tends to become local. The velocity of an isolated step can then acquire a universal form which depends on the local geometry, particularly the step curvature. 

To describe this situation for a step train, we applied asymptotics on a boundary integral formalism for the adatom fluxes. A highlight of our results is the emergence of the free boundary velocity as a linear superposition of equilibrium adatom densities in the same as well as adjacent steps. The contribution from the same step depends on the local curvature both kinetically, through the leading-order behavior of a double-layer potential, and thermodynamically via the step chemical potential. The kinetic interactions with neighboring steps, on the other hand, are expressed by decaying exponentials of effective terrace widths. In the language of boundary layer theory~\cite{Hinch-book}, these terms signify effects of the inner and outer regions associated with the step boundaries.

Our analysis explicitly yields effective kinetic lengths that enter the step velocity law. In particular,  step permeability causes the appearance of length scales that involve $\Ld$ and two convex-type combinations of lengths associated with the step edge barrier asymmetry. 

Our results motivate further studies of crystal growth in the step flow regime. For example, one can numerically compare the local geometric laws of our asymptotics to the morphological evolution of multiple, kinetically interacting steps in specific crystalline materials. \color{black} Aspects of the 2D step morphological evolution under strong anisotropy in vacuum are left unresolved. For some anisotropic step free energies, the formation of geometric singularities, e.g, micro-facets and corners, on steps poses a challenge.  Another direction of analytical interest concerns the effect of solvents on the step velocity, when the crystal surface is immersed in a liquid.

\medskip

\acknowledgments
 
The authors are indebted to C. Ratsch and J.~D. Weeks for valuable conversations. 
D.M. is grateful to R.~V. Kohn and A.~G. Shtukenberg for bringing useful bibliography to this author's attention. D.M. also wishes to thank Y. Giga for insightful comments on the theory of motion by curvature.
This work was partly supported  by the NSF Grant No. DMS-1412769 at the University of Maryland.

\appendix 

\section{On the case of concentric circular steps}
\label{app:radial}
In this appendix, we provide details for the radial case that were omitted in Sec.~\ref{subsec:explicit-radial}. We also outline how the approximation scheme of our integral formalism for non-circular steps (Sec.~\ref{sssec:scheme}) can properly reduce to the equations of motion for concentric circular steps when rotational symmetry holds (Sec.~\ref{subsec:explicit-radial}). In particular, we show that in the radial geometry the single- and double-layer potential terms are evaluated by use of modified Bessel functions. 

First, let us revisit the radial setting of Sec.~\ref{subsec:explicit-radial}. We will express the total adatom flux at a point of a circular step in a form that will later enable us to make direct comparisons to the boundary integral formalism. Consider the shifted adatom concentration on the $i$th terrace $\Omega^{(i)}$, viz.,
\begin{equation*}
	C_i(r)=a_i I_0(R)+b_i K_0(R)~,\ R=r/\Ld~,\ r_i<r<r_{i+1}~.
\end{equation*}
The values of the normal derivative of $C=C_i$ at the bounding steps, $\Gamma^{(i)}$ and $\Gamma^{(i+1)}$, are written in the matrix form
\begin{equation*}
    \begin{bmatrix}
        \left ( \frac{\partial C}{\partial \nu} \right )^+_i \\ 
        \left ( \frac{\partial C}{\partial \nu} \right )^-_{i+1}
    \end{bmatrix}
    = \Ld^{-1}\begin{bmatrix} 
    I_1(R_i) & -K_1(R_i) \\
    I_1(R_{i+1}) & -K_1(R_{i+1})
    \end{bmatrix}
    \begin{bmatrix} a_i \\ b_i
    \end{bmatrix}~,\ R_i=\frac{r_i}{\Ld}~,
\end{equation*}
where $(\partial C/\partial\nu)^\pm_j=\partial C/\partial r$ at $r=r_j$ with $j=i$ ($+$ sign) or $j=i+1$ ($-$ sign). On the other hand, the Robin-type conditions~\eqref{eq:bcf_bc_general} for the radial adatom flux  yield the system
\begin{align*} \mA_i \begin{bmatrix} a_i \\ b_i
\end{bmatrix}
=\begin{bmatrix}
    \ueq_i \\ \ueq_{i+1}
\end{bmatrix}~,
\end{align*}
where the matrices $\mA_i$ are defined by 
\begin{equation*} 
    \mA_i=
    \begin{bmatrix} 
        I_0(R_i)-\frac{\Lad^+}{\Ld} I_1(R_i)   & K_0(R_i)+\frac{\Lad^+}{\Ld} K_1(R_i)  \\
        I_0(R_{i+1})+\frac{\Lad^-}{\Ld} I_1(R_{i+1})  & K_0(R_{i+1})-\frac{\Lad^-}{\Ld} K_1(R_{i+1}) 
    \end{bmatrix}.
\end{equation*}
Evidently, $\det(\mA_i)=-\Lad^+\Lad^- \Lambda_i$; $\Lambda_i$ is defined in Sec.~\ref{subsec:explicit-radial}.

Thus, the total flux into step $\Gamma^{(i)}$ for $i \neq 1,N$ is 
\begin{align}
    J_{i}^{\rm tot} &= J_{i,\perp}^{\rm tot}=D_s  \left [ \left ( \frac{\partial C_i}{\partial \nu} \right )^+_i - \left ( \frac{\partial C_{i-1}}{\partial \nu} \right )^-_i \right ]  \nonumber \\
    =&\frac{D_s}{\Ld} \begin{bmatrix} I_1(R_i) \\ -K_1(R_i) \end{bmatrix}^T \left \{
     \mA_i^{-1} \begin{bmatrix} \ueq_i \\ \ueq_{i+1} \end{bmatrix} - \mA_{i-1}^{-1} \begin{bmatrix} \ueq_{i-1} \\ \ueq_i \end{bmatrix} \right \}, \label{eq:flux_from_classical_1}
\end{align}
where the superscript $T$ here denotes the transpose.
This result serves our purpose of connecting the explicit equations of motion for the radial case to our boundary integral formalism. In fact, we will show that Eq.~\eqref{eq:flux_from_classical_1} exactly agrees with the respective outcome of our boundary integral equations.

Next, we turn our attention to the formalism of Sec.~\ref{sec:main}, which we will place in the radial setting. We begin by substituting the Robin-type boundary conditions~\eqref{eq:bcf_adj_bc} 
directly into Eq.~\eqref{eq:Green-form}. It is more algebraically convenient to proceed this way, instead of invoking
$\lim_{\my \to \mx} \mtd_i[f](\my)=\mp f(\mx)/2 + \msd_i[f](\mx)$, 
because in this setting $\mtd_i[1](\my)$ has a simple expression, 
as we will see below. 

The result of the above substitution for the terrace $\Omega^{(i)}$ (with $C_i=C$) reads
\begin{align}
    & \left ( \mts_{i+1} + \Lad^- \mtd_{i+1} \right ) \left [ \left ( \frac{\partial C}{\partial \nu} \right )^- \right ](\mx) \nonumber \\ 
    & + \left ( -\mts_i + \Lad^+ \mtd_i \right ) \left [ \left ( \frac{\partial C}{\partial \nu} \right )^+ \right ](\mx) \nonumber \\
    & = \mtd_{i+1}[\ueq_{i+1}](\mx) - \mtd_i[\ueq_i](\mx) - C(\mx)~,
\end{align}
where $\mx$ lies in $\Omega^{(i)}$. This formalism can be extended to the extremal terraces, where $i=0$ or $i=N$, 
by the introduction of zero terms pertaining to the  
(nonexistent) steps $\Gamma^{(0)}$ and $\Gamma^{(N+1)}$. Accordingly, we can obtain a system of integral equations on curves $\Gamma^{(i)}$ and  $\Gamma^{(i+1)}$, which bound terrace $\Omega^{(i)}$, by allowing $\mx$ to approach each of these steps from inside $\Omega^{(i)}$. 

Now let us focus on simplifications due to the radial geometry. We notice that the (assumed isotropic) step free
energy, the fluxes, the layer potentials $\mts_j[1](\mx)$ and $\mtd_j[1](\mx)$, and the shifted density $\ueq_j$ are all
constant along the respective side of a given step edge. Consequently, we can now use  
the same replacement that was previously employed for the derivation of Eq.~\eqref{eqs:ies_first_approx}, bearing in mind that in the present setting this equation is exact. The result is the system
\begin{subequations} \label{eqs:appendix_step1}
    \begin{align}
        & \left ( \mts_{i+1}[1](\mx) + \Lad^- \mtd_{i+1}[1](\mx) \right ) \left ( \frac{\partial C}{\partial \nu} \right )^-(\mx) \nonumber \\ 
        & + \left \{ -\mts_i[1](\mx) + \lim_{\my \to \mx} \left (\Lad^+ \mtd_i[1](\my) \right ) \right \} \left ( \frac{\partial C}{\partial \nu} \right )^+(\mx) \nonumber \\
        & = \mtd_{i+1}[1](\mx) \ueq_{i+1}(\mx) - \lim_{\my \to \mx} \left ( \mtd_i[1](\my) \right ) \ueq_i(\mx) \nonumber \\
        &  -\ueq_i(\mx) - \Lad^+ \left ( \frac{\partial C}{\partial \nu} \right )^+(\mx)~,\  \mx\ \mbox{in}\ \Gamma^{(i)}~;
    \end{align}
    and
    \begin{align}
        & \left ( \mts_{i+1} [1](\mx) + \lim_{\my \to \mx} \left [ \Lad^- \mtd_{i+1}[1](\my) \right ] \right ) \left ( \frac{\partial C}{\partial \nu} \right )^-(\mx) \nonumber \\ 
        & + \left ( -\mts_i[1](\mx) + \Lad^+ \mtd_i[1](\mx) \left ( \frac{\partial C}{\partial \nu} \right )^+ \right ](\mx) \nonumber \\
        & = \lim_{\my \to \mx} \left [ \mtd_{i+1}[1](\my) \right ] \ueq_{i+1}(\mx) - \mtd_i[1](\mx) \ueq_i(\mx) \nonumber \\
        & -\ueq_{i+1}(\mx) + \Lad^- \left ( \frac{\partial C}{\partial \nu} \right )^-(\mx)~,\ \mx\ \mbox{in}\ \Gamma^{(i+1)}~.
    \end{align}
    \end{subequations}
 
 It remains for us to evaluate the requisite layer potentials when the steps $\Gamma^{(i)}$ are concentric circles with radii $r_i$. First, consider the single-layer potential terms. By using polar coordinates, for $\mx$ in $\Gamma^{(j)}$ we have 
    \begin{equation*}
       \mts_k[1](\mx) = -\frac{\Ld}{2\pi} \int_{-\pi}^\pi K_0 \left ( \sqrt{R_j^2 + R_k^2 - 2R_j R_k \cos\theta} \right ) R_k\, d \theta~.
    \end{equation*}
    Consider Graf's addition formula \cite{Bateman-II}, viz.,
    \begin{equation*}
        K_0 \left ( \sqrt{Z^2+\Xi^2-2Z\Xi\cos\theta} \right ) = \sum_{n=-\infty}^\infty K_n(Z) I_n(\Xi)\, e^{in\theta}~,
    \end{equation*}
    where $Z>0$, $\Xi \geq 0$ and $Z\geq \Xi$ (for real $Z$ and $\Xi$). By using this formula, and interchanging the order of integration and summation for $\mts_k[1](\mx)$, we find (for $\mx$ in $\Gamma^{(j)}$)
    \begin{align}\label{eq:appendix_singlelayer_result}
        & \mts_k[1](\mx)= -\frac{r_k}{2\pi}\nonumber \\
        & \times \sum_{n=-\infty}^\infty \int_{-\pi}^\pi K_n(\max \{ R_j,R_k \}) I_n(\min \{ R_j,R_k \}) e^{in\theta}\, d \theta \nonumber \\
        & = -\Ld R_k K_0(\max \{ R_j,R_k \}) I_0(\min \{ R_j,R_k \})~.
    \end{align}

  Subsequently, we proceed to evaluate the double-layer potential terms. We distinguish 
    the following cases. We start with the term $\mtd_{i+1}[1](\mx)$, where the evaluation point $\mx$ lies in terrace $\Omega^{(i)}$. In this case, we 
    need the normal derivative 
    \begin{align*}
        & \frac{\partial}{\partial \nu(\my)}G(\mx-\my) = -\frac{1}{2\pi} \frac{\partial}{\partial r_k} \sum_{n=-\infty}^\infty K_n(R_k) I_n \Biggl ( \frac{|\mx|}{\Ld} \Biggr) e^{in\theta} \nonumber \\
        & = \frac{1}{4\pi\Ld} \sum_{n=-\infty}^\infty \left \{K_{n+1}(R_k)+K_{n-1}(R_k) \right\} I_n \Biggl( \frac{|\mx|}{\Ld}  \Biggr) e^{in\theta}~, 
    \end{align*}
      where $|\my|=r_k>|\mx|$. Note that the derivative is taken with respect to the larger radius.
For $k=i+1$, we obtain 
\begin{equation*}
    \mtd_{i+1}[1](\mx) = R_{i+1} K_1(R_{i+1})\, I_0 \Biggl( \frac{|\mx|}{\Ld} \Biggr)~,\ \mx\ \mbox{in}\ \Omega^{(i)}~.
\end{equation*}
  
The next calculation concerns the double-layer potential term $\mtd_i[1](\mx)$, where $\mx$ lies in $\Omega^{(i)}$. The requisite normal derivative involves changing the smaller radius. We compute
\begin{align*}
    \frac{\partial}{\partial \nu(\my)}&[G(\mx-\my)]= -\frac{1}{4\pi\Ld} \nonumber \\ 
    & \times \sum_{n=-\infty}^\infty \left \{ I_{n+1}(R_k)+I_{n-1}(R_k) \right\} K_n \left ( \frac{|\mx|}{\Ld} \right) e^{in\theta}~,
\end{align*}
for $|\my|=R_k \Ld< |\mx|$.
Upon integration with $k=i$, we have
\begin{equation*}
    \mtd_i[1](\mx) = -R_i I_1(R_i)\, K_0 \Biggl( \frac{|\mx|}{\Ld} \Biggr)~,\ \mx\ \mbox{in}\ \Omega^{(i)}~.
\end{equation*}

At this stage, we comment on the
relation between $\mtd_i(\my)$ and $\mtd_i(\mz)$ when $\my$ lies in $\Omega^{(i)}$ and $\mz$ lies in $\Omega^{(i-1)}$, and both points approach point $\mx$ in step $\Gamma^{(i)}$. We explicitly compute
\begin{align} \label{eq:comp-lim-D}
    & \lim_{\my \to \mx} \mtd_i[1](\my) - \lim_{\mz \to \mx} \mtd_{i}[1](\mz) \notag\\
    & = R_i \{K_1(R_i) I_0(R_i)+K_0(R_i) I_1(R_i)\}= 1 
\end{align}
where $R_i=|\mx|/\Ld$. 
The above formula is a special case of the limit $\lim_{\my \to \mx} \mtd_i[f](\my)=\mp f(\mx)/2 + \msd_i [f](\mx)$, 
used in Sec.~\ref{sec:main}. Equation~\eqref{eq:comp-lim-D} comes from a Wronskian~\cite{Bateman-II} and is useful throughout our algebraic manipulations. 

With the expressions for the layer potentials at hand, we can rewrite integral equation system~\eqref{eqs:appendix_step1}. Thus, we obtain
\begin{subequations} \label{eqs:appendix_formalism}
    \begin{align}
        & R_{i+1}\left \{ -\Ld K_0(R_{i+1}) I_0(R_i) + \Lad^- K_1(R_{i+1}) I_0(R_i) \right\} \left ( \frac{\partial C}{\partial \nu} \right )^- \nonumber \\ 
        & + R_i\left \{ \Ld K_0(R_i) I_0(R_i) - \Lad^+ I_1(R_i) K_0(R_i) \right\} \left ( \frac{\partial C}{\partial \nu} \right )^+ \nonumber \\
        & = R_{i+1} K_1(R_{i+1}) I_0(R_i) \ueq_{i+1} + R_i K_1(R_i) I_0(R_i) \ueq_i \nonumber \\
        & \qquad -\ueq_i - \Lad^+ \left ( \frac{\partial C}{\partial \nu} \right )^+~,
    \end{align} 
    \begin{align}
        & R_{i+1}\left\{ -\Ld K_0(R_{i+1}) I_0(R_{i+1}) + \Lad^- K_1(R_{i+1}) I_0(R_{i+1}) \right\} \left ( \frac{\partial C}{\partial \nu} \right )^-\nonumber \\ 
        & + R_i\left\{ \Ld K_0(R_{i+1}) I_0(R_i) - \Lad^+ I_1(R_i) K_0(R_{i+1}) \right\} \left ( \frac{\partial C}{\partial \nu} \right )^+ \nonumber \\
        & = R_{i+1} K_1(R_{i+1}) I_0(R_{i+1}) \ueq_{i+1} + R_i K_1(R_{i+1}) I_0(R_i) \ueq_i \nonumber \\
        & \qquad -\ueq_{i+1} + \Lad^- \left ( \frac{\partial C}{\partial \nu} \right )^-~.
    \end{align} 
 \end{subequations}
Here, $C=C_i$ and $(\partial C/\partial\nu)^\pm$ denotes the normal derivative of $C$ evaluated at point $\mx$ of step $\Gamma^{(i)}$ ($+$ sign) or point $\my$  of step $\Gamma^{(i+1)}$ ($-$ sign).
 By solving system~\eqref{eqs:appendix_formalism}, we realize that the total flux into step $\Gamma^{(i)}$ takes the form
\begin{align}
        J_i^{\rm tot} &= D_s \left \{ \begin{bmatrix} 1 \\ 0 \end{bmatrix}^T
        \mB_i^{-1} \mD_i \begin{bmatrix} \ueq_i \\ \ueq_{i+1} \end{bmatrix} - \begin{bmatrix} 0\\ 1 \end{bmatrix}^T \mB_{i-1}^{-1} \mD_{i-1} \begin{bmatrix} \ueq_{i-1} \\ \ueq_i \end{bmatrix} \right\}. \label{eq:flux_from_IF_1}
    \end{align}
The matrices $\mB_i$ and $\mD_i$ are defined by
 \begin{widetext}
    \begin{equation*}
    \mB_i=\begin{bmatrix} R_i\{\Lad^+ I_0(R_i) K_1(R_i) + \Ld K_0(R_i) I_0(R_i)\}
        & R_{i+1}\{\Lad^- K_1(R_{i+1}) I_0(R_i) - \Ld K_0(R_{i+1}) I_0(R_i)\} \\
        R_i\{-\Lad^+ I_1(R_i) K_0(R_{i+1}) + \Ld K_0(R_{i+1}) I_0(R_i)\}
        & -R_{i+1}\{\Lad^- K_0(R_{i+1}) I_1(R_{i+1})+\Ld K_0(R_{i+1}) I_0(R_{i+1})\}
    \end{bmatrix}~,
\end{equation*}
\end{widetext}
\begin{equation*}
    \mD_i=\begin{bmatrix}
        -R_i I_0(R_i) K_1(R_i) & R_{i+1} K_1(R_{i+1}) I_0(R_i) \\
        R_i I_1(R_i) K_0(R_{i+1}) & -R_{i+1} K_0(R_{i+1}) I_1(R_{i+1})
    \end{bmatrix}.
\end{equation*}

Let us summarize the results of this appendix so far.  On the one hand, the formalism of Sec.~\ref{subsec:explicit-radial} yields Eq.~\eqref{eq:flux_from_classical_1} for $J_i^{\rm tot}$ in terms of $\ueq_j$ ($j=i,\,i\pm 1$), by use of matrices $\mA_i$. On the other hand, the boundary integral equation formalism of Sec.~\ref{sec:main}, with the evaluation of the layer potentials via Graf's addition formula, furnishes Eq.~\eqref{eq:flux_from_IF_1}. It remains to compare Eqs.~\eqref{eq:flux_from_classical_1} and~\eqref{eq:flux_from_IF_1}. To this end, we carry out some algebra in which we apply Eq.~\eqref{eq:comp-lim-D}. We omit further details here. 
The explicit calculation of $J_i^{\rm tot}$ by each formula, Eq.~\eqref{eq:flux_from_classical_1} and Eq.~\eqref{eq:flux_from_IF_1},  yields  the same expression, viz., 
\begin{widetext}
    \begin{align}
        J_i^{\rm tot} =& D_s \left \{
        -\left ( \frac{\left[ \Lad^+ K_1(R_{i-1}) + \Ld K_0(R_{i-1}) \right] I_1(R_i) - \left [\Lad^+ I_1(R_{i-1}) - \Ld I_0(R_{i-1}) \right] K_1(R_i)}{\tilde\Lambda_{i-1}} \right. \right. \nonumber \\
        & + \left. \frac{\left [\Lad^- I_1(R_{i+1}) + \Ld I_0(R_{i+1}) \right] K_1(R_i) - \left [\Lad^- K_1(R_{i+1}) - \Ld K_0(R_{i+1}) \right] I_1(R_i)}{\tilde \Lambda_i} \right ) 
        \ueq_i \nonumber \\
        & \left. + \frac{\Ld}{R_i}\left(\frac{\ueq_{i+1}}{\tilde\Lambda_i}  
        + \frac{\ueq_{i-1}}{\tilde\Lambda_{i-1}}
        \right) \right\}\qquad (i\neq 1,\,N)~;\qquad \tilde\Lambda_i=-\Ld^2\det(\mA_i)~. \label{eq:appendix_flux_final}
    \end{align}
\end{widetext}
Recall that $\tilde\Lambda_i$ is related to the $\Lambda_i$ introduced in Sec.~\ref{subsec:explicit-radial} by $\tilde\Lambda_i=\Ld^2\Lad^+\Lad^-\Lambda_i$.


\nocite{*}
\bibliography{references}

\begin{thebibliography}{69}
\bibitem [{Not()}]{Note-facets}%
  \BibitemOpen
  \href@noop {} {}\bibinfo {note} {Our analysis should hold for steps
  represented by non-self-intersecting curves that are twice continuously
  differentiable. To avoid technicalities, we simply state that the step curves
  are smooth, i.e., infinitely differentiable.}\BibitemShut {Stop}%
\bibitem [{Com()}]{Comment-sign}%
  \BibitemOpen
  \href@noop {} {}\bibinfo {note} {The two different signs on the right-hand
  side of this limit are due to our particular definition of the unit normal
  vector $\nv$ along the boundary of the $i$th terrace, $\Omega^{(i)}$.
  Specifically, $\nv$ points toward (outward from) this terrace on step
  $\Gamma^{(i)}$ ($\Gamma^{(i+1)}$).}\BibitemShut {Stop}%
\end{thebibliography}

\end{document}